\documentclass[12pt]{article}
\usepackage{amsmath,amssymb}
\usepackage{graphicx}

\usepackage[affil-it]{authblk}

\usepackage[left=2cm,right=2cm,top=2cm,bottom=2cm,a4paper]{geometry}

\usepackage{hyperref}

\hypersetup{
pdfstartview=FitV,
colorlinks=true,
linkcolor=blue, 
citecolor=red, 
filecolor=black, 
urlcolor=blue}

\def\beq{\begin{eqnarray}}
\def\eeq{\end{eqnarray}}

\allowdisplaybreaks

\begin{document}

\setcounter{footnote}{0}
\setcounter{figure}{0}
\setcounter{table}{0}

\title{\textbf{Muon $g-2$ in Anomaly Mediated SUSY Breaking}}

\author{Debtosh Chowdhury\thanks{debtosh.chowdhury@roma1.infn.it}}
\author{Norimi Yokozaki\thanks{norimi.yokozaki@roma1.infn.it}}

\affil{\small Istituto Nazionale di Fisica Nucleare, Sezione di Roma, \authorcr {\it Piazzale Aldo Moro 2, I-00185 Rome, Italy}}

\date{}

\maketitle

\begin{abstract}
\noindent
Motivated by two experimental facts, the muon $g-2$ anomaly and the observed Higgs boson mass around 125 GeV,  we propose a simple model of anomaly mediation, which can be seen as a generalization of mixed modulus-anomaly mediation. In our model, the discrepancy of the muon $g-2$ and the Higgs boson mass around 125 GeV are easily accommodated. The required mass splitting between the strongly and weakly interacting SUSY particles are naturally achieved by the contribution from anomaly mediation. This model is easily consistent with $SU(5)$ or $SO(10)$ grand  unified theory.
\end{abstract}

\clearpage

\section{Introduction}

The anomalous magnetic moment of the muon, the muon $g-2$, has been measured very precisely at the Brookhaven E821 experiment~\cite{gm2_exp}:
\begin{eqnarray}
(a_\mu)_{\rm exp} =(116\,592\, 08.9 \pm 6.3) \times 10^{-10}.
\end{eqnarray}
Notably, $(a_\mu)_{\rm exp}$  deviates  from standard model (SM) predictions beyond $3\sigma$ level. The deviation, $\Delta a_{\mu} \equiv (a_{\mu})_{\rm exp} - (a_{\mu})_{\rm SM} $, is known to be
\begin{eqnarray}
\Delta a_\mu = 
\left\{
\begin{array}{cc}
(26.1 \pm 8.0)\times 10^{-10} &   \cite{gm2_hagiwara}    \\
(28.7 \pm 8.0)\times 10^{-10} &   \cite{gm2_davier}   
\end{array}
\right\},
\end{eqnarray}
where $(a_{\mu})_{\rm SM}$ is the SM prediction. Since the size of $(\Delta a_\mu)$ is comparable to that of the electroweak contribution in the SM~\cite{gm2_sm_ew}, 
a plausible possibility is that new particles with masses of $\mathcal{O}(100)$ GeV are responsible for $(\Delta a_\mu)$:
the anomaly of the muon $g-2$ may be a clear evidence that physics beyond SM exists around the weak scale.

In the minimal supersymmetic standard model (MSSM), the discrepancy of the muon $g-2$ is explained if the smuons,  chargino and neutralino  are as light as $\mathcal{O}(100)$ GeV with $\tan\beta=\mathcal{O}(10)$~\cite{gm2_susy1, gm2_susy2}. 
Also, supersymmetry (SUSY) provides us with attractive features in addition to the explanation for the muon $g-2$ anomaly: a solution to the hierarchy problem and a framework for the grand unified theory (GUT). Therefore, to consider SUSY models explaining $\Delta a_\mu$ is one of the important directions for physics beyond the SM.

However, there is an obstacle in this direction. 
The squarks and gluino have not yet been observed at the Large Hadron Collider (LHC), resulting in the lower bound on their mass at 1.4-1.8 TeV~\cite{lhc_susy}.
 Moreover, the observed Higgs boson mass $m_h$ around $125\, {\rm GeV}$~\cite{lhc_higgs} can be explained, only if there is  a sizably large radiative correction from the heavy stop(s)~\cite{higgs_susy}, unless the large trilinear coupling of the stops exists. 
In fact, including higher order corrections beyond the 3-loop level, it is suggested that the stop is as heavy as 3-5 TeV~\cite{higgs_3loop} in the absence of the large trilinear coupling of the stops.  Since squarks and sleptons belong to a same representation of $SU(5)$ GUT gauge group and the gaugino masses unify at the high energy scale in a simple setup,  it is rather nontrivial to obtain the heavy stop and light sleptons simultaneously. As a consequence, to construct a convincing SUSY scenario for the muon $g-2$ is a rather difficult task.

Recently there has been  a resurgence of interest in explaining both the muon $g-2$ anomaly and the observed Higgs boson mass within a unified framework.
It has been shown that the discrepancy of the muon $g-2$ and the Higgs boson mass around 125 GeV can be explained simultaneously by introducing GUT breaking effects,\footnote{
In Refs.~\cite{split_gm2_1, split_gm2_2}, $\Delta a_\mu$ and the Higgs boson mass around 125 GeV are successfully explained  without introducing a GUT breaking effect on the soft SUSY breaking masses. The models shown in Refs are based on the ``Split-Family SUSY", where  the third generation sfermions are much heavier than the first and second generation sfermions. Also, extensions of the MSSM allow us to explain $\Delta a_\mu$ without introducing a GUT breaking effect (see e.g. Ref. \cite{vector}).
} in the gauge mediation~\cite{gm2_gauge}, gaugino mediation~\cite{gm2_gaugino, gm2_gaugino2},\footnote{
The models shown in Refs. \cite{gm2_gaugino} are attractive, since they are free from the SUSY and strong CP problem as well as  the SUSY flavor problem.  
Non-universal gaugino masses are naturally obtained based on the product group unification model, which solves the  notorious doublet-triplet splitting problem~\cite{DT_split, DT_split2}.
}
and gravity mediation~\cite{gm2_gravity}. In most of these cases, the violation of the GUT relation among gaugino masses is at least required.

In this paper, we show that the required mass splitting among the strongly and weakly interacting SUSY particles, i.e. the GUT breaking effect on the soft SUSY breaking masses, is naturally induced from anomaly mediation~\cite{amsb1, amsb2}\footnote{
While completing this manuscript, Ref.~\cite{note_added} appeared in arXiv, 
which has some similarity in the starting point.
}
: both the Higgs boson mass around 125 GeV and $\Delta a_\mu$ can be easily explained in our simple framework, which is consistent with $SO(10)$ or $SU(5)$ GUT.

The rest of the paper is organized as follows: in section \ref{sec:pAMSB} we propose the phenomenological AMSB (pAMSB) model used in our analysis. In section \ref{sec:mhgm2} we discuss the SUSY contribution to the muon $g-2$ in our setup and show numerical results. A more fundamental realization of the pAMSB model is shown in section \ref{sec:pamsbr}. 
Finally,  section \ref{sec:conc} is devoted to the conclusion and discussion.


\section{Phenomenological AMSB Model} \label{sec:pAMSB}

In SUSY models, masses of squarks and sleptons are required to be highly split in order to explain the Higgs boson mass around 125 GeV and the muon $g-2$ anomaly simultaneously. Moreover, the bino and wino masses should be (much) smaller than the gluino mass at the high energy scale, otherwise the radiative corrections lift up the slepton masses and it becomes difficult to accommodate the experimental result of the muon $g-2$. 

The anomaly mediated SUSY breaking (AMSB) contributes to the masses of the colored and non-colored SUSY particles very differently: the squark and gluino masses obtain large contributions, while the slepton, bino and wino get negative or small contributions. This feature of AMSB is welcome for the Higgs mass around 125 GeV and the muon $g-2$.
Based on this observation, we propose a phenomenological AMSB (pAMSB) model, which can be easily accommodated into $SU(10)$ or $SU(5)$ grand unified theory.

Within a supergravity framework, we construct the pAMSB model with the following K\"{a}hler potential:
\beq
K = -3 M_P^2 \ln \left[1- \frac{f_{\rm hid}}{3 M_P^2} - \frac{Q_{\rm SM}^\dag Q_{\rm SM}}{3M_P^2} - \frac{\Delta f}{3M_P^2} \right],
\eeq
where $f_{\rm hid}$ is a function of hidden sector superfields, and $Q_{\rm SM}$ is a chiral superfield in the MSSM. The reduced Planck mass is denoted by $M_P$ ($M_P \simeq 2.4 \cdot 10^{18}$\,GeV).
The superpotential is also assumed to be separated as $W=W_{\rm vis} +W_{\rm hid}$, where $W_{\rm vis}$ and $W_{\rm hid}$ are superpotentials for the visible sector and hidden sector superfields, respectively. (A concrete example of $f_{\rm hid}$ and $W_{\rm hid}$ is shown in Appendix \ref{sec:mirage-like}.) Here, $\Delta f$ is an additional source of the sfermion masses, and is defined later. 
In the case $\Delta f =0$, the K\"{a}hler potential is so-called sequestered form and the scalar masses vanish at the tree level. Scalar masses (gaugino masses) are generated at the two-loop  level (one-loop level) from anomaly mediation (see Appendix A.1). 
The squark and slepton masses are estimated as
\beq
{m}_{Q_i}^{\prime \, 2}(2\, {\rm TeV}) &=&  \left[8.40 -2.27\, \delta_{i 3} \right] M_0^2, \nonumber \\
m_{\bar{U}_i}^{\prime \,2}(2\, {\rm TeV}) &=&  \left[ 8.50 -3.81\, \delta_{i 3}\right] M_0^2,  \nonumber \\
m_{{\bar{D}}_i}^{\prime \,2}(2\, {\rm TeV}) &=& \left[8.62   -0.72\, \delta_{i 3} \right] M_0^2,  \nonumber \\
m_{{{L}}_i}^{\prime \,2}(2\, {\rm TeV}) &=&  \left[-0.34 -0.05\, \delta_{i 3}  \right] M_0^2,\nonumber \\
m_{{\bar{E}}_i}^{\prime \,2}(2\, {\rm TeV}) &=& \left[ -0.37 -0.10\,  \delta_{i 3} \right] M_0^2, \label{eq:pure_amsb}
%
%
\eeq
where $Q_i$, $\bar U_i$ and $\bar D_i$ denote a left-handed quark, right-handed up-type quark and right-handed down-type quark, and $L_i$ and $\bar E_i$ are left-handed lepton and right-handed lepton, respectively. The index $i$ represents a generation of a chiral multiplet.
The common mass scale from anomaly mediation is denoted by $M_0=m_{3/2}/(16\pi^2)$, where $m_{3/2}$ is the gravitino mass.
We evaluate the above soft masses at 2 TeV for $\tan\beta=20$, $m_t({\rm pole})=173.34$ GeV and $\alpha_s(M_Z)=0.1185$. 
The first term (second term) in the bracket comes from the gauge (Yukawa) interactions.
The corrections from 1st and 2nd generation Yukawa couplings are neglected.
Using one-loop beta-functions of gauge couplings, the gaugino masses are~\footnote{
The signs of $A_{klm}$ and $M_a$ have been flipped by the $R$-rotation: $A_{klm} \to e^{2i \theta_R} A_{klm}$ and $M_a \to e^{2 i \theta_R} M_a$. The definition of the $A$-term is given by $V \ni A_{klm} y_{klm} Q_k Q_l Q_m + h.c.$
}
\beq
M_{1}(2\, {\rm TeV}) = 1.43 M_0 , \ M_{2}(2\, {\rm TeV}) = 0.41 M_0 , \ M_{3}(2\, {\rm TeV}) = -3.12 M_0 , \label{eq:amsb_gaugino}
\eeq
where $M_1$, $M_2$ and $M_3$ are the masses of the bino, wino and gluino, respectively: $M_1:M_2:M_3 \simeq 7 :2:-15$.

We see that from Eqs.(\ref{eq:pure_amsb}) and (\ref{eq:amsb_gaugino}) the masses of strongly interacting SUSY particles ($M_3$, $m_{Q}'$, $m_{\bar U}'$) and weakly interacting ones ($M_2$, $m_{L}'$, $m_{\bar E}'$) are highly split and it may be useful for explaining the muon $g-2$ anomaly and the Higgs boson masses simultaneously. However, the slepton masses $m'_{L}$ and $m'_{\bar{E}}$ are tachyonic, since it interacts only non-asymptotically free gauge interactions.

The tachyonic sleptons can be avoided if there is an additional source of the scalar masses, contained in $\Delta f$:
\beq
\Delta f  &=&   -\frac{(x -\left<x\right>)^2}{2\left<x\right>^2} \Bigr[ c_{10}  (Q^\dag Q + \bar{U}^\dag \bar{U} + \bar{E}^\dag \bar{E})  \nonumber \\ 
&& + c_{\bar 5} (L^\dag L + \bar{D}^\dag \bar{D}) + c_{H_u} H_u^\dag H_u + c_{H_d} H_d^\dag H_d \Bigl] \nonumber \\
&& - \left[ d_{H_u} \frac{x-\left<x\right>}{\left<x\right>} \right] H_u^\dag H_u 
 - \left[ d_{H_d} \frac{ x-\left<x\right>}{\left<x\right>}\right] H_d^\dag H_d, \label{eq:pamsb_setup}
\eeq
where $H_u$ and $H_d$ are up-type and down-type Higgs, respectively. Here, $x=X+X^\dag$, and $X$ is a moduli field which has a non-zero $F$-term $F_X$: ${\left<F_X\right>}/{\left<x\right>} = \mathcal{O}(m_{3/2}/100)$.
The above type of $\Delta f$ with the suppressed $F$-term, $F_X$, arises if $X$ couples to the matter fields. 
Note that $\left<F_X\right>/\left<x\right>\sim (m_{3/2}/100)$  is obtained with a KKLT-type superpotential~\cite{kklt} (see also  Appendix \ref{sec:mirage-like}). 
The moduli $X$ in $\Delta f$ gives corrections to the soft SUSY breaking masses of the MSSM fields comparable to those from anomaly mediation. These corrections uplift the tachyonic slepton masses.
The setup in Eq.(\ref{eq:pamsb_setup}) is similar to that of  the {mixed modulus-anomaly mediation} scenario~\cite{mirage1, endo}, but allowing non-universal contributions to the soft masses from the moduli $X$. 

Moreover, unlike the mixed modulus-anomaly mediation, we can independently chose the soft masses squared and the trilinear coupling of the stops $A_t$ determined by $d_{H_u}$: large contributions to soft masses squared from $X$ do not always lead to large $A$-terms.
This  significantly enlarges the parameter space for explaining the muon $g-2$ anomaly and the Higgs boson mass around 125 GeV simultaneously, especially in cases that the Higgsino mass term $\mu$ is small (see discussion in Sec.~\ref{sec:pamsbr}).
Note that $\Delta f$ is consistent with $SU(5)$ GUT, and  it is also consistent with $SO(10)$ GUT if $c_5=c_{10}$. 

With $\Delta f \neq 0$, the scalar masses are modified from Eq.(\ref{eq:pure_amsb}). The scalar masses including $\Delta f$ are given by
\beq
m_{(Q, \, \bar{U}, \, \bar{E})}^2 &=& {m}_{(Q, \, \bar{U}, \, \bar{E})}^{\prime \,2} + m_{10}^2 , \nonumber \\
m_{Q_3}^2 &=& {m}_{Q_3}^{\prime \,2} + m_{10}^2   + m_{Q_3, \rm mixed}^2, \nonumber \\
m_{\bar{U}_3}^2 &=& {m}_{\bar{U}_3}^{\prime \,2} + m_{10}^2 + m_{\bar{U}_3, \rm mixed}^2, \nonumber \\
m_{(L, \, \bar{D})}^2 &=& {m}_{(L, \, \bar{D})}^{\prime \,2} + m_{\bar 5}^2 , \nonumber \\
m_{H_u}^2 &=& {m}_{H_u}^{\prime \,2} +  \delta m_{H_u}^2   + m_{H_u, \rm mixed}^2, \nonumber \\
m_{H_d}^2 &=& {m}_{H_d}^{\prime \,2} + \delta m_{H_d}^2 .  \label{eq:additional0}
\eeq
where $m^2_{{\bar 5},10}=c_{{\bar 5}, 10} |\left<F_X\right>/\left<x\right>|^2$, $\delta m_{H_d}^2 = c_{H_d} |\left<F_X\right>/\left<x\right>|^2$, and $\delta m_{H_u}^2 = (c_{H_u} +d_{H_u}^{\,2}) |\left<F_X\right>/\left<x\right>|^2$. 
All the parameters are defined at the GUT scale ($\sim 10^{16}$ GeV), that is, 
a mass from anomaly mediation $m'_k$ ($k \in [Q_i, \bar{U}_i, \bar{E}_i, L_i, \bar D_i, H_u, H_d]$) is evaluated using the gauge and Yukawa couplings at the GUT scale.
For simplicity, we set $d_{H_d}=0$ here and hereafter.
The trilinear coupling of stops and the mixed mass terms are 
\begin{eqnarray}
 \delta A_t &=& d_{H_u} {\left<F_X\right>}/{\left<x\right>}, \nonumber \\
m_{H_u, \rm mixed}^2 &=&  - 3Y_t^2 (\delta A_t + h.c.) M_0 ,  \nonumber \\ 
m_{Q_3, \rm mixed}^2 &=&  - Y_t^2 (\delta A_t + h.c.) M_0,  \nonumber \\
m_{\bar U_3, \rm mixed}^2 &=&  - 2Y_t^2 (\delta A_t + h.c.) M_0 . \label{eq:mixed1}
\end{eqnarray}

The gaugino masses can be also modified by introducing couplings between field strength superfields of vector multiplets and $X$. The gauge kinetic functions are 
\beq
\mathcal{L} \ni  \frac{1}{4} \int d^2\theta \left[\frac{1}{g_a^2} + 2 c _\lambda \frac{(X-\left<X\right>)}{\left<X\right>}\right]W_{\alpha}^a W^{\alpha\,a}  + h.c.
\eeq
Then the gaugino masses get an additional contribution as
\beq
M_{a} =  \delta M_{1/2} + \frac{\beta_a}{g_a} (16\pi^2 M_0),
\eeq
where $ \delta M_{1/2}\sim (m_{3/2}/100)$ and $\beta_a$ is the beta-function of the  gauge coupling $g_a$: an additional contribution to the gaugino masses comparable to those from anomaly mediation can arise.
The scalar masses are modified from Eq.(\ref{eq:additional0}) as
\begin{eqnarray}
m_k^2 \to m_k^2 + (m_k^2)_{\rm mixed},
\end{eqnarray}
where 
\begin{eqnarray}
(m_k^2)_{\rm mixed} &=& -\frac{1}{2}(\delta M_{1/2}  + h.c.) g_a^2 \frac{\partial \gamma_k}{\partial g_a^2} \, m_{3/2} \nonumber \\
&=& -\frac{1}{2}(\delta M_{1/2}  + h.c.) g_a^2 (4 C_a(k)) \, \frac{m_{3/2}}{16\pi^2}.
\label{eq:mixed2}
\end{eqnarray}
Here, $\gamma_k$ is the anomalous dimension of the superfield $k$, $\gamma_k = (\partial \ln Z_k)/(\partial \ln \mu)$ and $C_a(k)$ is a quadratic Casimir invariant of the field $k$ ($C_1(k)=(3/5)Q_{Y_k}^2$).

So far, the SUSY breaking masses at the GUT scale in pAMSB are summarized as follows:
\beq
M_{a} = \delta M_{1/2} + \frac{\beta_a}{g_a} (16\pi^2 M_0),
\eeq
\beq
A_{t} = -\frac{\beta_{Y_t}}{Y_t} (16\pi^2 M_0) + \delta A_t, \ \, A_{b} = -\frac{\beta_{Y_b}}{Y_b} (16\pi^2 M_0), \ \, A_{\tau} = -\frac{\beta_{Y_{\tau}}}{Y_{\tau}} (16\pi^2 M_0), 
\eeq
\beq
m_{(Q, \, \bar{U}, \, \bar{E})}^2 &=& {m}_{(Q, \, \bar{U}, \, \bar{E})}^{\prime \,2} + m_{10}^2  + (m_{(Q, \, \bar{U}, \, \bar{E})}^2)_{\rm mixed} , \nonumber \\
m_{(L, \, \bar{D})}^2 &=& {m}_{(L, \, \bar{D})}^{\prime \,2} + m_{\bar 5}^2
+ (m_{(L, \, \bar{D}}^2)_{\rm mixed}, \nonumber \\
m_{H_u}^2 &=& {m}_{H_u}^{\prime \,2} + \delta m_{H_u}^2 
+ (m_{H_u}^2)_{\rm mixed}, \nonumber \\
m_{H_d}^2 &=& {m}_{H_d}^{\prime \,2} + \delta  m_{H_d}^2 
+ (m_{H_d}^2)_{\rm mixed} , \label{eq:additional}
\eeq
where $(m_k^2)_{\rm mixed}$ is a sum of the contributions from Eqs.(\ref{eq:mixed1}) and (\ref{eq:mixed2}), and $\beta_{Y_t}$, $\beta_{Y_b}$ and $\beta_{Y_\tau}$ are the beta-functions of the Yukawa couplings, $Y_t$, $Y_b$ and $Y_{\tau}$, respectively.
The soft SUSY breaking masses are written in terms of the following set of the parameters, 
\begin{eqnarray}
[M_0 (\equiv m_{3/2}/16 \pi^2\,), m_{10}^2, m_{\bar 5}^2, \delta m_{H_u}^2, \delta m_{H_d}^2, \delta M_{1/2}, \delta A_t].
\end{eqnarray}
In the limit $m_{10}^2=m_{\bar 5}^2=\delta m_{H_u}^2=\delta m_{H_d}^2$ and $\delta A_t=\delta M_{1/2}=0$, the mass spectrum of the SUSY particles corresponds to that of the minimal AMSB~\cite{mamsb}.\footnote{
See Refs.~\cite{mamsb_2} for phenomenological aspects of the minimal AMSB, where the SUSY contribution to the muon $g-2$ is also discussed. 
Also, in Ref.~\cite{amsb_other}, the phenomenological aspects of anomaly mediation models are considered without imposing the muon $g-2$ constraint.
}

\section{Muon $g-2$ in the pAMSB} \label{sec:mhgm2}

In this section, we check whether the muon $g-2$ anomaly and the observed Higgs boson mass around 125 GeV can be explained in the pAMSB model. 
The SUSY contribution to the muon $g-2$, $(\delta a_\mu)_{\rm SUSY}$, is sufficiently large in the following three cases:
\begin{enumerate}
\item[ (a) ]  The wino, Higgsino and muon sneutrino are light.
\item[ (b) ] The bino and left-handed smuon as well as the right-handed smuon are light.
\item[ (c) ] The intermediate case between (a) and (b).
\end{enumerate}

In the first case (a), the wino-Higgsino-(muon sneutrino) loop dominates $(\delta a_\mu)_{\rm SUSY}$. 
This contribution is estimated as~\cite{gm2_susy2}
\begin{eqnarray}
(\delta a_\mu)_{\tilde{W}-{\tilde H}-{\tilde \nu}} &\simeq& (1- \delta_{2L}) \frac{\alpha_2}{4\pi}\frac{m_\mu^2 M_2 \mu}{m_{\tilde \nu}^4}\tan\beta \cdot 
F_C \left(\frac{\mu^2}{m_{\tilde \nu}^2}, \frac{M_2^2}{m_{{\tilde \nu}}^2} \right), \nonumber \\
&\simeq& 18.2 \times 10^{-10} \left(\frac{500 \, {\rm GeV}}{m_{\tilde \nu}}\right)^2 \frac{\tan\beta}{25}, \label{eq:wino_gm2}
\end{eqnarray}
where $m_{\tilde \nu}$ is the mass of the muon sneutrino, and we take $\mu=(1/2) m_{\tilde \nu}$ and $M_2 = m_{\tilde \nu}$ in the second line.
The soft mass parameters as well as $\mu$ in the R.H.S. of Eq.(\ref{eq:wino_gm2}) are defined at the soft mass scale. 
A leading two-loop correction from large QED-logarithms is denoted by $\delta_{2L}$, which is given by~\cite{photonic, photonic_recent}
\begin{eqnarray}
\delta_{2L} =  \frac{4 \alpha}{\pi} \ln \frac{m_{\tilde \nu} }{m_\mu}.
\end{eqnarray}
To explain $\Delta a_\mu = (26.1 \pm 8.0)\cdot 10^{-10}$ by $(\delta a_\mu)_{\tilde{W}-{\tilde H}-{\tilde \nu}}$, the masses of the wino and the muon sneutrino should be smaller than around 500 GeV.

In the second case (b), the $\tilde{B}-{\tilde \mu}_L-{\tilde \mu}_R$ diagram dominates $(\delta a_\mu)_{\rm SUSY}$. 
The $\tilde{B}-{\tilde \mu}_L-{\tilde \mu}_R$ contribution is found to be~\cite{gm2_susy2}
\begin{eqnarray}
(\delta a_\mu)_{\tilde{B}-{\tilde \mu}_L-{\tilde \mu}_R} &\simeq& 
(1- \delta_{2L}) \frac{3}{5}\frac{\alpha_1}{4\pi}\frac{m_\mu^2 \mu}{M_1^3}\tan\beta \cdot 
F_N \left(\frac{m_{{\tilde \mu}_L}^2}{M_1^2}, \frac{m_{{\tilde \mu}_R}^2}{M_1^2} \right), \nonumber \\
&\simeq& 21.7 \times 10^{-10} \frac{\mu}{3200\, {\rm GeV}} \frac{\tan\beta}{8} \left(\frac{110\, {\rm GeV}}{M_1}\right)^3, \label{eq:gm2_bino}
\end{eqnarray}
where we take $m_{\tilde{\mu}_L}=3 M_1$ and $m_{\tilde{\mu}_R}=2 M_1$ in the second line. One can see that a very light bino with a mass $\sim 100$ GeV is required to explain the muon $g-2$ anomaly.

Note that we do not need to consider the case $(c)$. This is because the light bino and wino can not be obtained  simultaneously.
The bino and wino mass at 2\,TeV are 
\begin{eqnarray}
M_{1}(2\,{\rm TeV})&=& 0.43 \,\delta M_{1/2} + 1.43 M_0, \nonumber \\
M_{2}(2\,{\rm TeV}) &=&0.82  \,\delta M_{1/2} + 0.41 M_0, \label{eq:wino_bino}
\end{eqnarray}
at the one-loop level.
In the case the bino mass is small, say, $M_{1}(2\,{\rm TeV}) \simeq 0.2 M_0$, the additional contribution to the gaugino masses is $\delta M_{1/2} = -2.9 M_0$; however, the wino mass becomes $M_{2}(2\,{\rm TeV}) \simeq -2.0 M_0$, and hence, it is impossible to obtain the light bino and wino simultaneously. Because of this reason, we have only two possibilities (a) and (b) to explain $\Delta a_\mu$.

\subsection{Small $\mu$ case} \label{seq:small_mu}
First, we consider the small $\mu$ case with $\delta M_{1/2} = 0$. 
In this case, the gaugino masses are  same as those in anomaly mediation. 
As shown in Eq.(\ref{eq:amsb_gaugino}), the wino is the lightest gaugino, and it is expected that $(\delta a_\mu)_{\tilde{W}-{\tilde H}-{\tilde \nu}}$ is enhanced if $\mu$ is small. On the other hand, 
it is difficult to enhance $(\delta a_\mu)_{\tilde{B}-{\tilde \mu}_L-{\tilde \mu}_R}$ because of the large bino mass. Therefore we concentrate on the wino-Higgsino-(muon sneutrino) contribution.

In our numerical calculation, the SUSY mass spectrum is calculated using {\tt Suspect 2.43}~\cite{suspect} with a modification suitable for our purpose. The Higgs boson mass ($m_h$) as well as the SUSY contribution to the muon $g-2$ ($(\delta a_\mu)_{\rm SUSY}$) is evaluated using {\tt FeynHiggs 2.10.4}~\cite{feynhiggs}. In the region where both Higgsino and wino are light, the branching ratio of ${\rm Br}(b\to s \gamma)$ is enhanced due to the SUSY contribution. We demand that the SUSY contribution do not exceed $2\sigma$ bound:
\begin{eqnarray}
-5.7\cdot 10^{-5}  < \Delta  {\rm Br}(b \to s \gamma) < 7.1 \cdot 10^{-5}, \label{eq:bsg}
\end{eqnarray}
where $\Delta {\rm Br}(b \to s \gamma) \equiv {\rm Br}(b \to s \gamma)_{\rm MSSM}-{\rm Br}(b \to s \gamma)_{\rm SM}$.
Here, we use the SM prediction in Ref.~\cite{bsg_sm} and the experimental value in Ref.~\cite{bsg_hfag}. 
We use {\tt SuperIso} package~\cite{superiso} to calculate $\Delta {\rm Br}(b \to s \gamma)$. 
Note that the constraint from ${\rm Br}(B_{s} \to \mu^+ \mu^-)$~\cite{bsmm} is not stringent in the parameter space of our interest, since the CP-odd Higgs boson mass $m_A$ is rather large.

\begin{figure}[!t]
\begin{flushleft}
\includegraphics[scale=0.425]{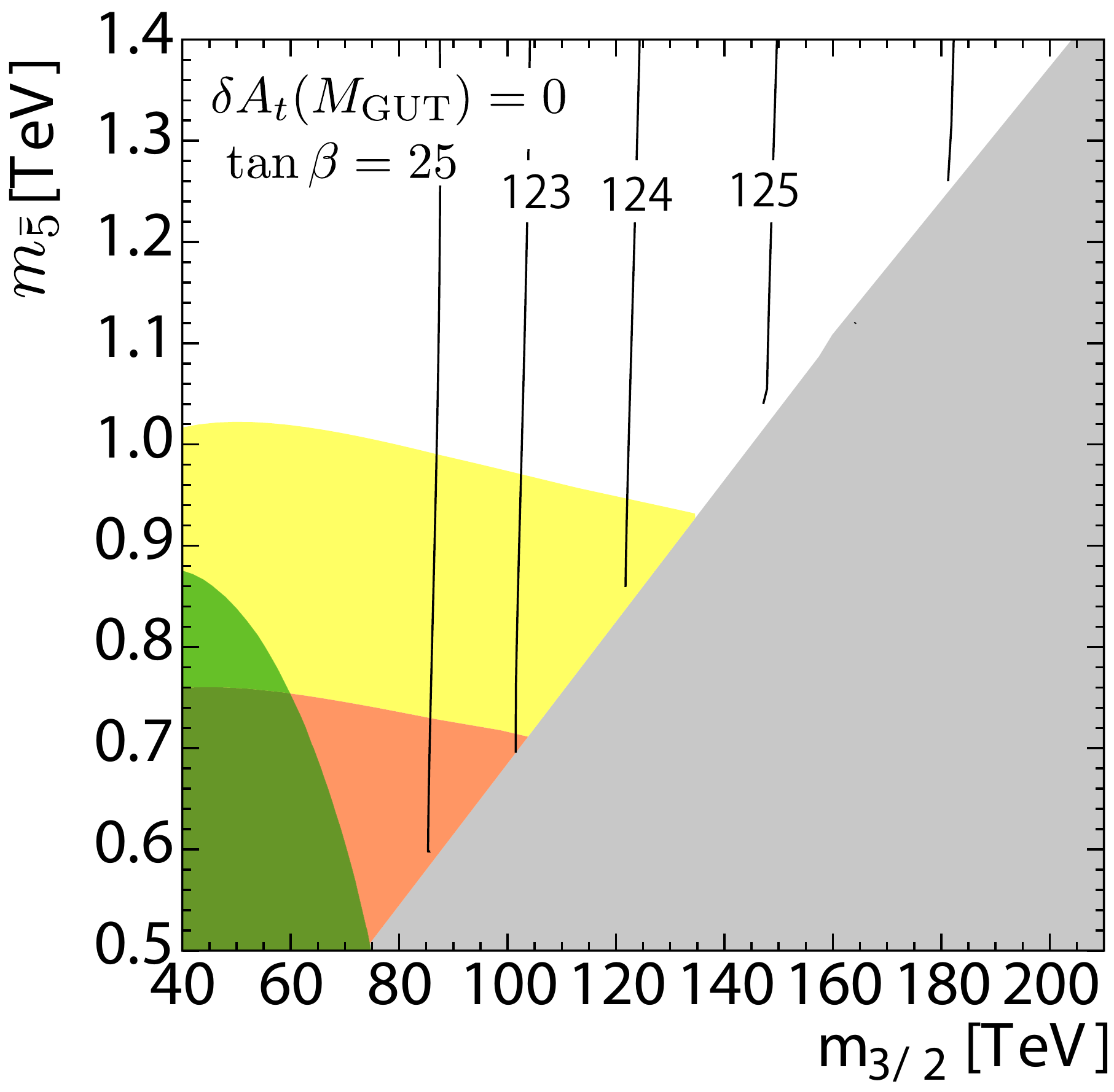}
\includegraphics[scale=0.425]{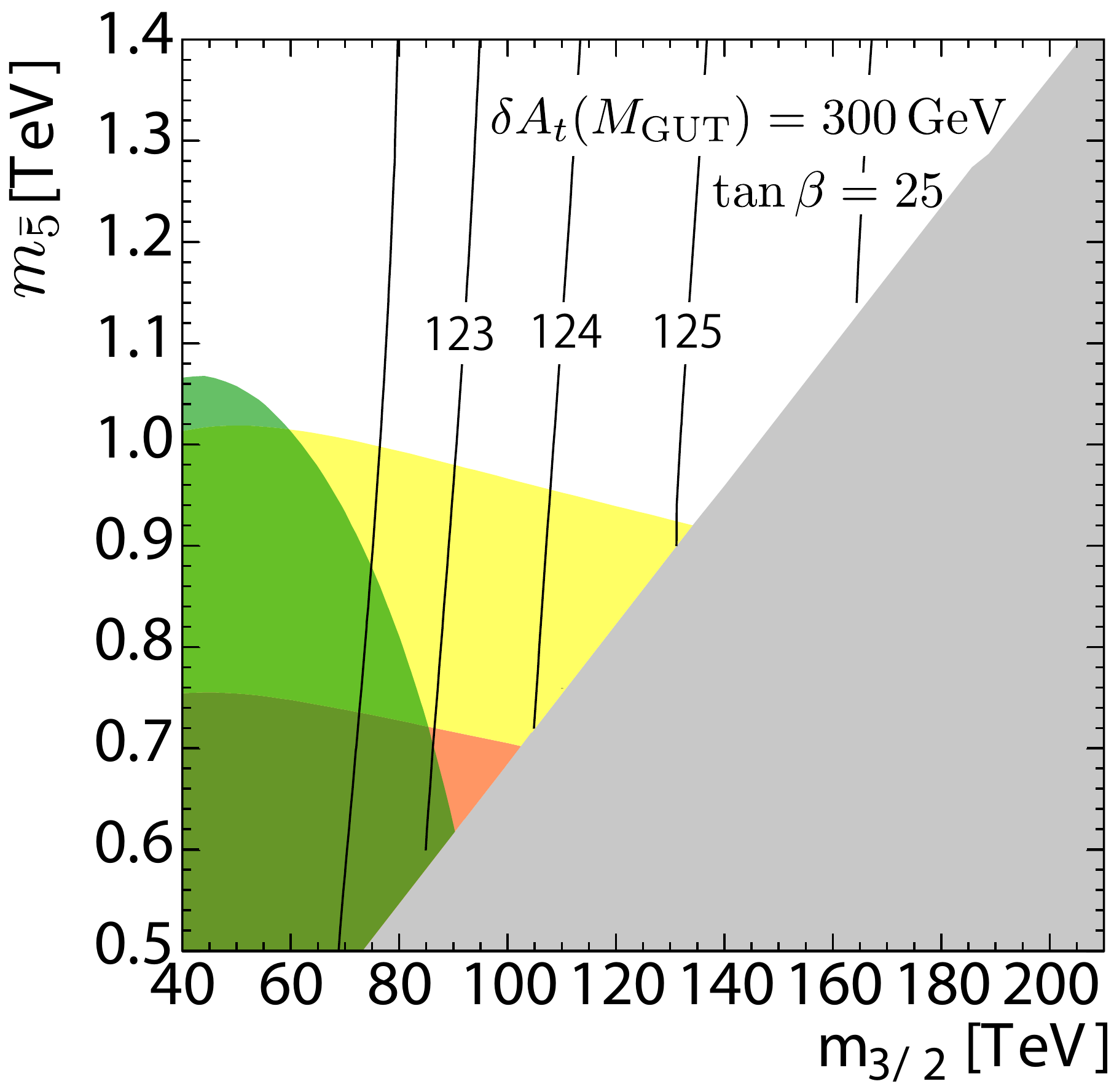}
\includegraphics[scale=0.425]{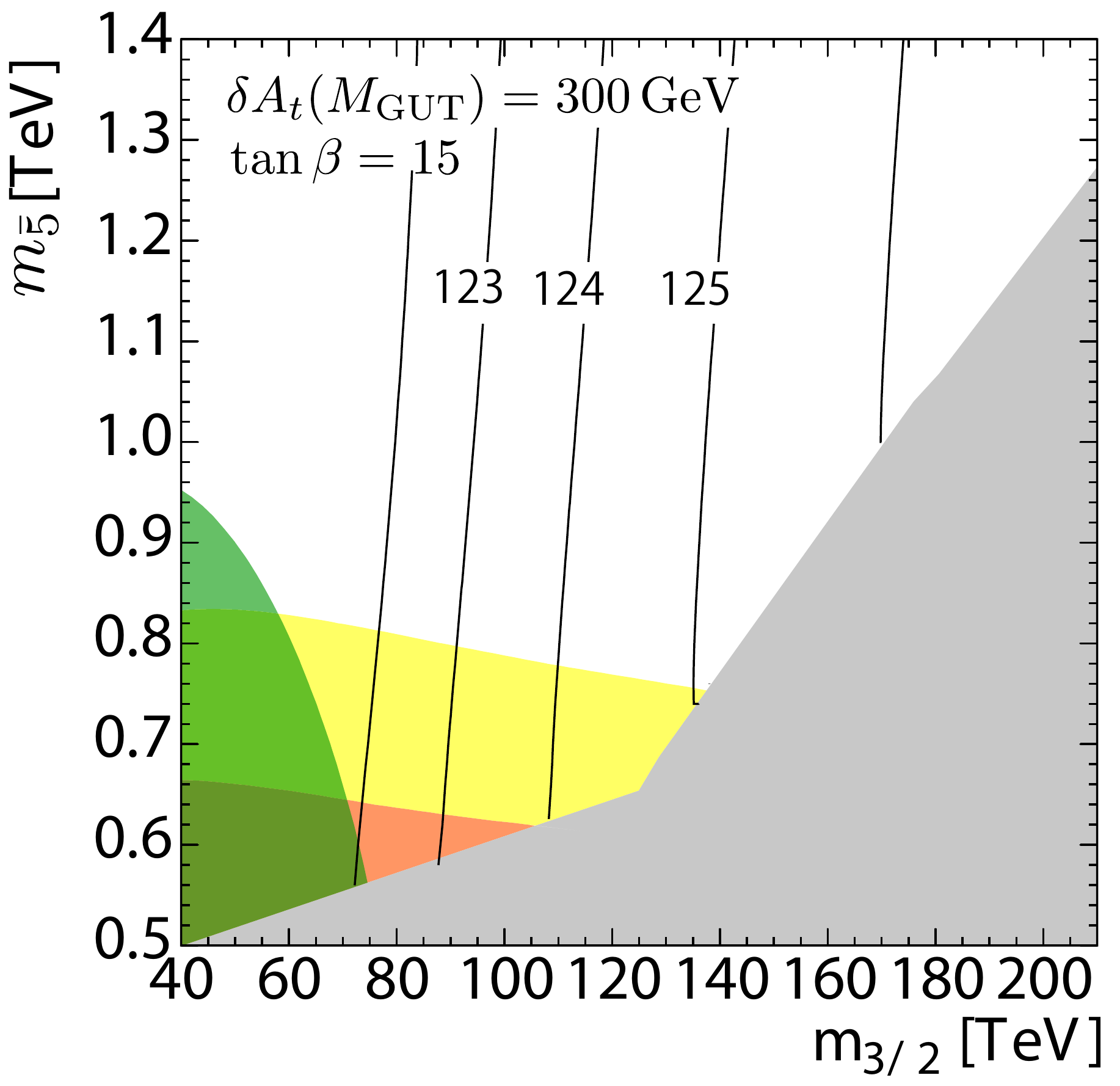}
\includegraphics[scale=0.425]{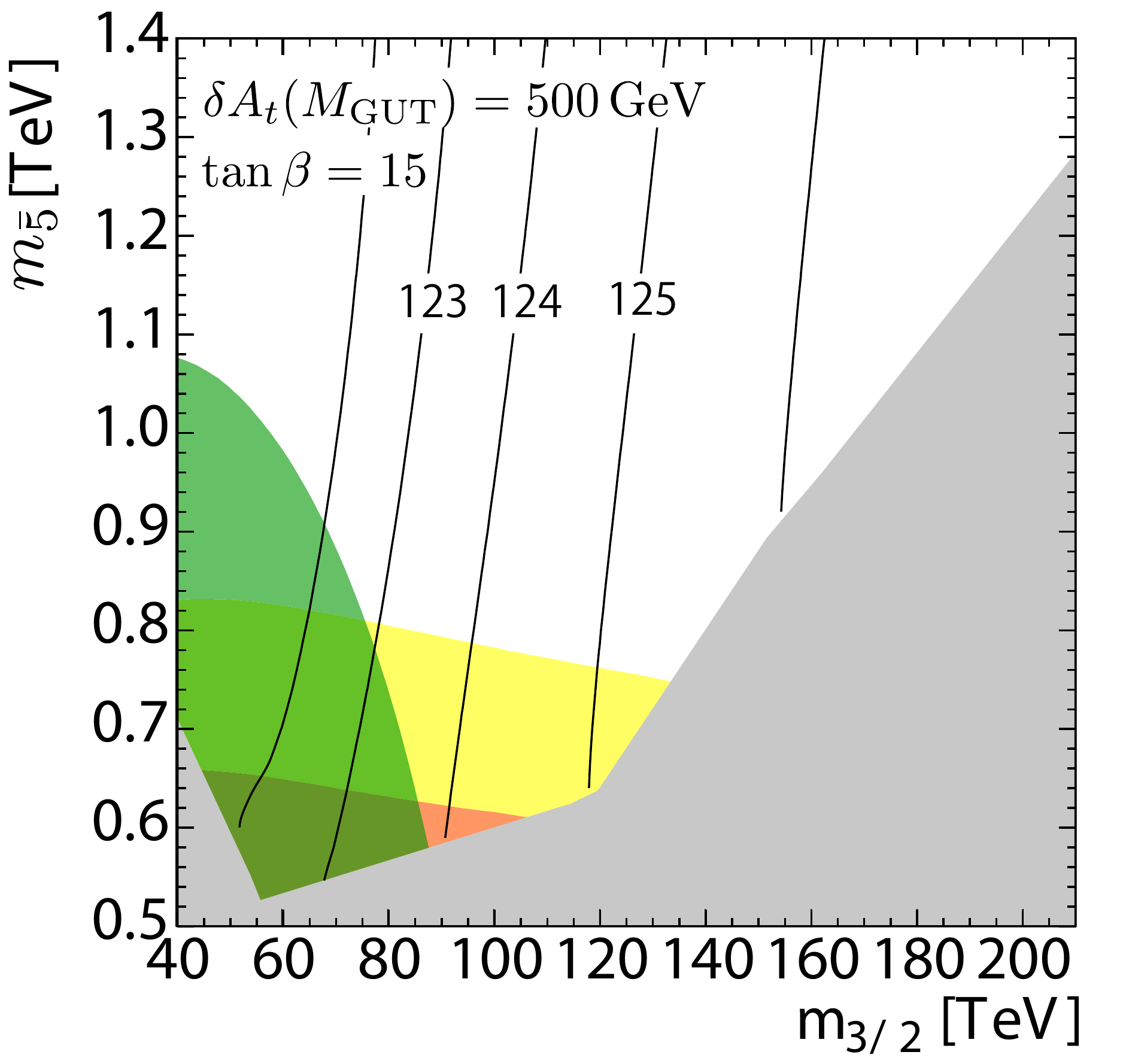}
\caption{The contours of $m_h$ (in the unit of GeV) and $(\delta a_\mu)_{\rm SUSY}$ for $m_{\bar 5}=m_{10}$. In these plots, $\delta M_{1/2}=0$ and $\mu=150$\,GeV. We take $m_A=1500$\,GeV ($m_A=2500$\,GeV) and $\tan\beta=25$ ($\tan\beta=15$) in the upper (lower) two panels. 
In the orange (yellow) region, the discrepancy of the muon $g-2$ is  reduced to 1$\sigma$ (2$\sigma$) level. 
In the green region, $\Delta {\rm Br}(b \to s \gamma)$ exceeds the $2\sigma$ bound.
Here, $m_t({\rm pole})=173.34$ GeV and $\alpha_s(m_Z)=0.1185$. 
}
\label{fig:smallmu1}
\end{flushleft}
\end{figure}

In Fig.~\ref{fig:smallmu1}, we plot the contours of $m_h$ and the region consistent with $\Delta a_\mu$.
We take $m_{10} = m_{\bar 5}$, which is consistent with $SO(10)$ GUT. 
We set $\mu=150$ GeV, $m_A=1500$ GeV and $\tan\beta=25$ ($\mu=150$ GeV, $m_A=2500$ GeV and $\tan\beta=15$) in the upper (lower) two panels. (The weak scale values of $\mu$ and $m_A$ are taken as input parameters instead of $\delta m_{H_u}^2$ and $\delta m_{H_d}$.)
Here, $m_t({\rm pole})=173.34$ GeV and $\alpha_s(m_Z)=0.1185$. In the orange (yellow) region, the discrepancy of the muon $g-2$ from the SM prediction is reduced to 1$\sigma$ $(2\sigma)$ level. The gray region is excluded due to the stop LSP (left-bottom) or stau LSP (right). In the green region, $\Delta {\rm Br}(b \to s \gamma)$ exceeds the 2$\sigma$ bound in Eq.~(\ref{eq:bsg}). The constraint from ${\rm Br}(b\to s \gamma)$ is rather severe and the region with large $\delta A_t$ is excluded. 
Note that one can not cancel between the chargino contribution and the charged Higgs contribution to ${\rm Br}(b \to s \gamma)$ by taking smaller $m_A$, since the both contributions are constructive to the SM value for $A_t, \mu>0$ at the soft mass scale. 
Still, as one can see the discrepancy of the muon $g-2$ can be reduced to $1 \sigma$ level. The calculated Higgs boson mass $m_h$ is consistent with the observed value around 125 GeV. 

Combined CMS and ATLAS measurement of Higgs mass allow a range from 124.6 to 125.6 GeV at 2$\sigma$~\cite{lhc_higgs}. On top of it the experimental uncertainty in the top mass measurement~\cite{top_mass_th} and theoretical uncertainty estimated by {\tt FeynHiggs 2.10.4} allow for at least $\pm 3 $\,GeV uncertainty in the Higgs boson mass value. Thus in all the plots we show the Higgs boson mass in the range 122-126 GeV.

\begin{figure}[!t]
\begin{flushleft}
\includegraphics[scale=0.43]{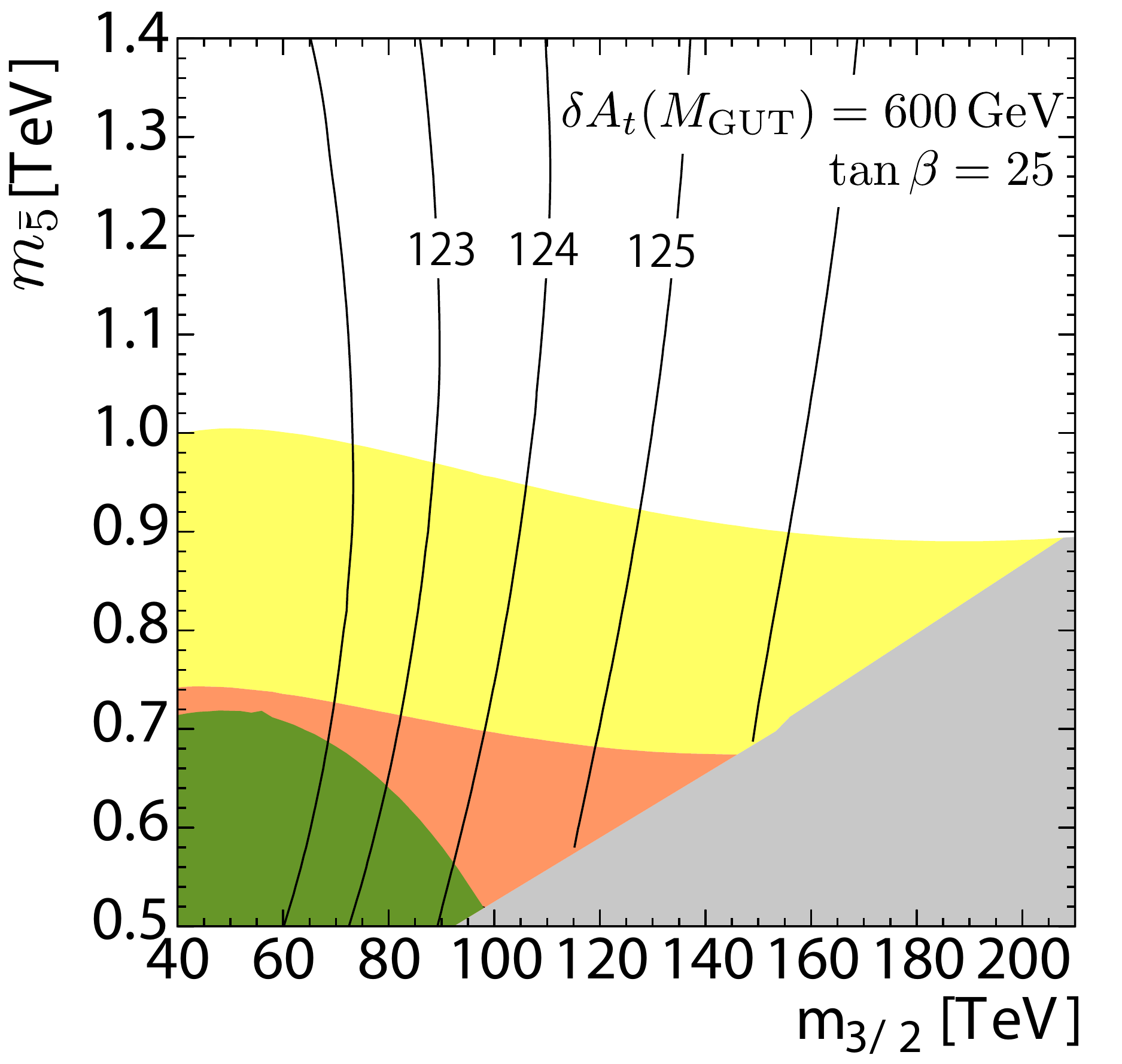}
\includegraphics[scale=0.43]{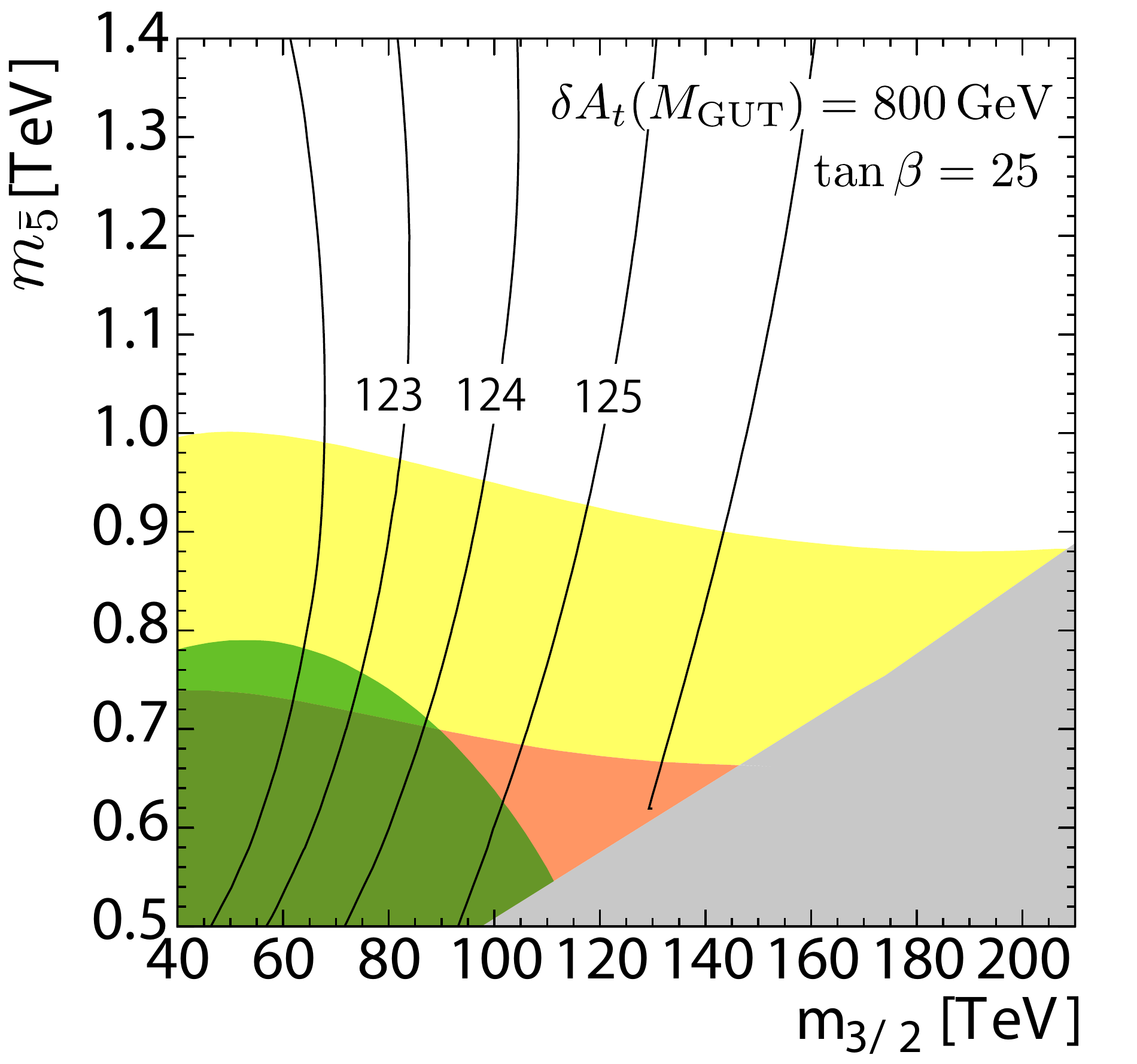}
\caption{The contours of $m_h$ (in the unit of GeV) and $(\delta a_\mu)_{\rm SUSY}$ for $m_{10}=\sqrt{3}m_{\bar 5}$. In the left (right) panel, $\delta A_t(M_{\rm GUT})=600$\,(800)\,GeV. Here, $\mu=150$\,GeV, $m_A=1500$\,GeV and $\tan\beta=25$.
The other parameters are same as in Fig.~\ref{fig:smallmu1}.
}
\label{fig:smallmu2}
\end{flushleft}
\end{figure}

\vspace{10pt}
Next, we relax the condition $m_{10}=m_{\bar 5}$. In this case, the muon $g-2$ anomaly and the Higgs boson mass around 125 GeV are more easily explained. We show the contours of $m_h$ and the region consistent with $\Delta a_\mu$ in Fig.~\ref{fig:smallmu2} for $m_{10}=\sqrt{3} m_{\bar 5}$. Because the heavier stops are allowed (($\bar{U}_3$, $Q_3$)$\in$ {\bf 10} in SU(5) GUT gauge group), the constraint from $\Delta {\rm Br}(b \to s \gamma)$ becomes less sever than the previous case with $m_{10}=m_{\bar 5}$. Moreover, the right-handed stau can be heavier and the region with tachyonic stau is reduced. As a result, the region which can explain the muon $g-2$ anomaly and the observed Higgs boson mass simultaneously becomes wider.

Also, we show sample mass spectra of different model points in Table \ref{table:spectrum1}. {\bf P1} ({\bf P2})
is consistent with $SO(10)$ ($SU(5)$) GUT, where $m_{10}/m_{\bar 5}=1.0\, (\sqrt{2})$ is taken. In both of the model points, the calculated Higgs boson mass $m_{h}$ is consistent with the observed value, and the discrepancy of the muon $g-2$ from the SM prediction is reduced to 1$\sigma$ level.
%
%
The squark masses as well as the gluino mass in {\bf P1}\,({\bf P2}) are around 2\,(3) TeV, and hence, it is expected that the squarks and gluino are discovered or excluded at the LHC with $\sqrt{s}=14$ TeV~\cite{lhc_future}.
The lightest neutralino is Higgsino-like mixed with the wino, therefore the relic abundance of this neutralino is too small to explain the observed dark matter abundance: we need another dark matter candidate, e.g. axion in the small $\mu$ cases.\footnote{
Although one can consider the non-thermal production~\cite{ntp_dm} (see also \cite{ntp_dm2}) of the lightest neutralino to explain the observed dark matter abundance, the neutralino-nucleon scattering cross section is too large; therefore, this possibility is excluded. 
}

\vspace{10pt}
Note that the existence of the small $\delta M_{1/2}$ is also helpful in the small $\mu$ case: it enlarges the parameter space which can explain the muon $g-2$ anomaly. This is because the small mass of the wino can always be obtained by choosing $\delta M_{1/2}$, regardless of the gravitino mass (see Eq.(\ref{eq:wino_bino})).

\vspace{10pt}
If one takes the bino mass to be small with $\delta M_{1/2} \ne 0$, the wino mass becomes large (see Eq.(\ref{eq:wino_bino})). Then, $(\delta a_\mu)_{\tilde{W}-{\tilde H}-{\tilde \nu}}$ is suppressed. In this case, we need $(\delta a_\mu)_{\tilde{B}-{\tilde \mu}_L-{\tilde \mu}_R} \gtrsim 1.8\cdot 10^{-9}$ to explain the muon $g-2$ anomaly. Since $(\delta a_\mu)_{\tilde{B}-{\tilde \mu}_L-{\tilde \mu}_R}$ is proportional to $\mu \tan\beta$ and large $\tan\beta$ easily leads to tachynic staus via radiative corrections,  we consider the case with large $\mu$ and moderate $\tan\beta$ for this purpose.

%
%
%
%

\begin{table}[t!]
    \caption{The mass spectra for small $\mu$ cases. We take $\delta M_{1/2}=0$, $\alpha_s(M_Z)=0.1185$ and $m_t({\rm pole})=173.34$ GeV.
   }
  \begin{center}
    \begin{tabular}{  c | c  }
            {\bf P1} & \\
\hline
    $m_{3/2}$ & 100 TeV \\
    $m_{\bar 5}$ & 700 GeV \\
    $\delta A_t(M_{\rm GUT})$ & 400 GeV\\
    $m_{10}$ & $m_{\bar 5}$\\
    $\tan \beta$ & 25 \\
    $\mu$ & 140 GeV \\
    $m_A$ & 1500 GeV \\
    \hline
\hline    
    $m_{\rm gluino}$ & 1.9 TeV \\
      $m_{\tilde{q}}$ & 2.0 TeV \\
    $m_{\tilde{t}_{1,2}}$ & 1.0, 1.5 TeV \\
    $m_{\tilde{e}_L} (m_{\tilde{\mu}_L})$ & 612 GeV\\
    $m_{\tilde{e}_R} (m_{\tilde{\mu}_R})$ & 482 GeV\\
    $m_{\tilde{\tau}_1}$ & 132 GeV\\
     $m_{\chi_1^0}$, $m_{\chi_2^0}$ & 126, 150 GeV \\
     $m_{\chi_3^0}$, $m_{\chi_4^0}$ & 351, 928 GeV \\
     $m_{\chi_1^{\pm}}$, $m_{\chi_2^{\pm}}$ & 133, 352 GeV \\
     $m_{h}$ & 124.1 GeV \\
    $ (\delta a_\mu)_{\rm SUSY}$ & $1.82 \cdot 10^{-9}$ \\
     $ \Delta {\rm Br}(b \to s \gamma)$ & $6.4 \cdot 10^{-5}$\\  	
    \end{tabular}
        \hspace{20pt}
        \begin{tabular}{  c | c  }
                    {\bf P2} & \\
\hline
    $m_{3/2}$ & 130 TeV \\
    $m_{\bar 5}$ & 650 GeV \\
    $\delta A_t(M_{\rm GUT})$ & 400 GeV\\
    $m_{10}$ & $\sqrt{2} m_{\bar 5}$ \\
    $\tan \beta$ & 25 \\
    $\mu$ & 150 GeV \\
    $m_A$ & 1500 GeV \\
    \hline
\hline    
    $m_{\rm gluino}$ & 2.8 TeV \\
      $m_{\tilde{q}}$ & 2.9 TeV \\
    $m_{\tilde{t}_{1,2}}$ & 1.6, 2.2 TeV \\
    $m_{\tilde{e}_L} (m_{\tilde{\mu}_L})$ & 665 GeV\\
    $m_{\tilde{e}_R} (m_{\tilde{\mu}_R})$ & 760 GeV\\
    $m_{\tilde{\tau}_1}$ & 239 GeV\\
     $m_{\chi_1^0}$, $m_{\chi_2^0}$ & 141, 159 GeV \\
     $m_{\chi_3^0}$, $m_{\chi_4^0}$ & 443, 1208 GeV \\
     $m_{\chi_1^{\pm}}$, $m_{\chi_2^{\pm}}$ & 147, 443 GeV \\
     $m_{h}$ & 125.1 GeV \\
    $ (\delta a_\mu)_{\rm SUSY}$ & $2.02 \cdot  10^{-9}$ \\
    $ \Delta {\rm Br}(b \to s \gamma)$ & $3.9 \cdot  10^{-5}$\\  	
    \end{tabular}
  \label{table:spectrum1}
  \end{center}
\end{table}

%
%
%
\subsection{Large $\mu$ case}
\begin{figure}[!t]
\begin{center}
\includegraphics[scale=0.42]{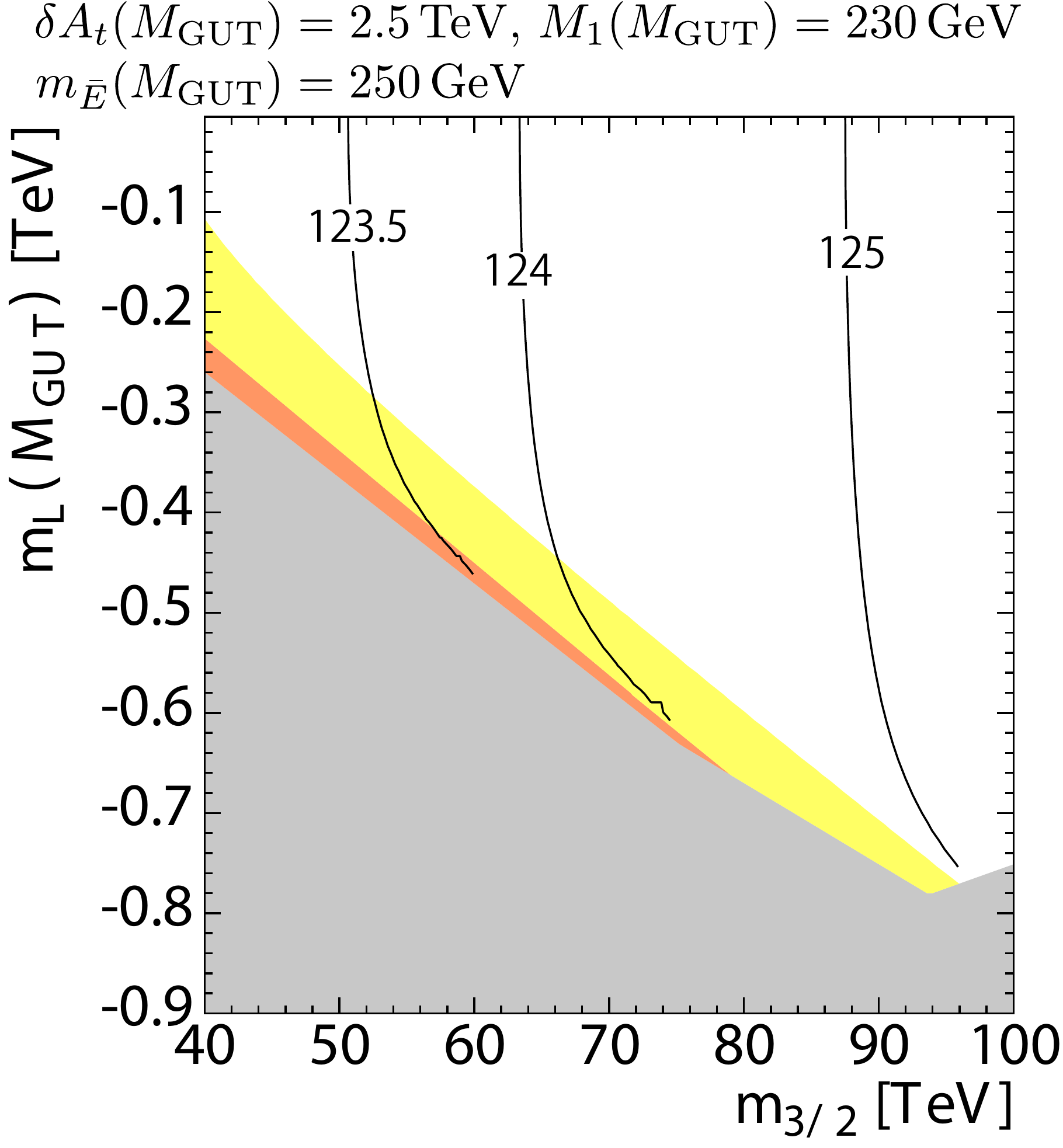}
\includegraphics[scale=0.43]{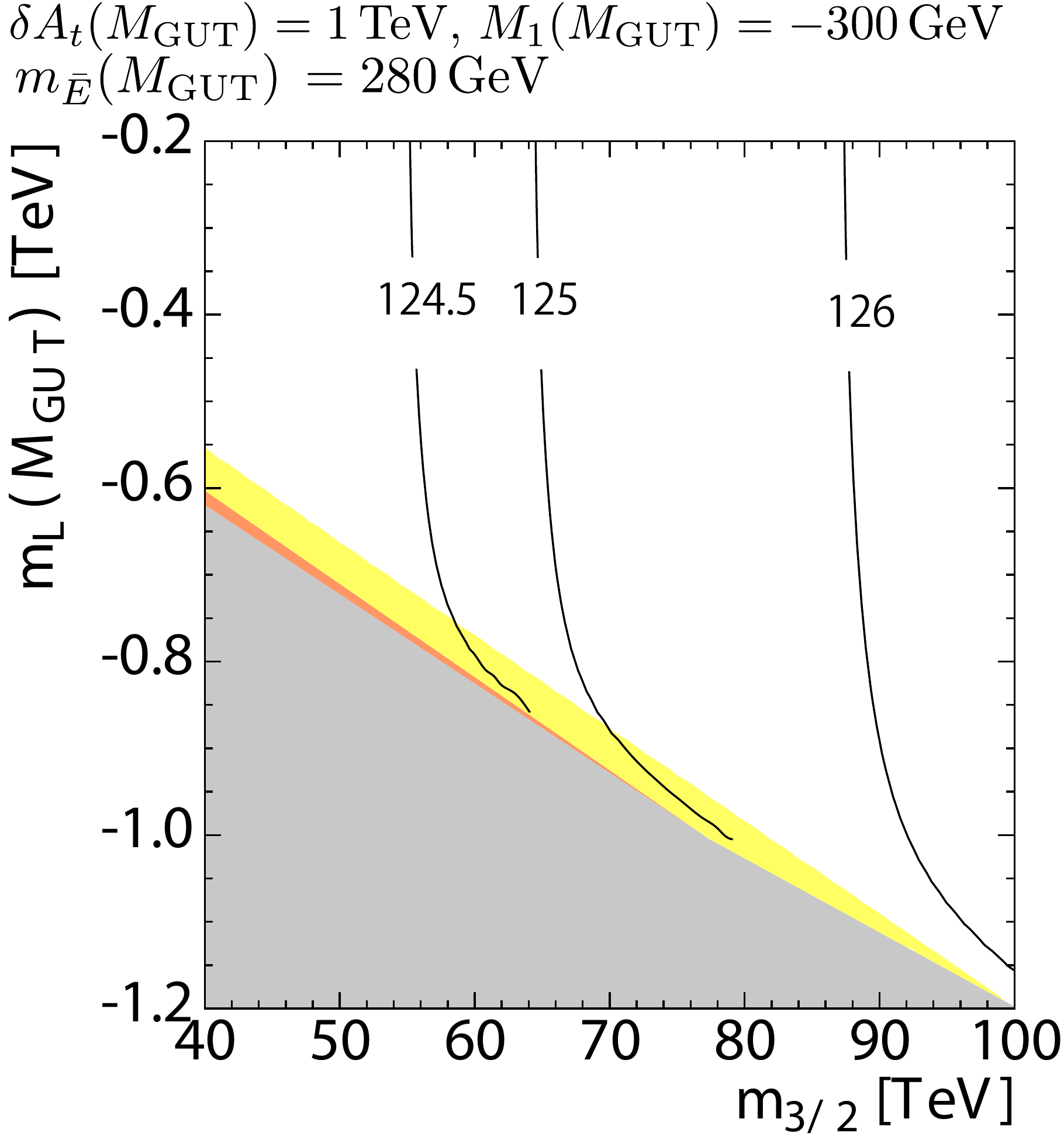}
\caption{The contours of $m_h$ (in the unit of GeV) and $(\delta a_\mu)_{\rm SUSY}$ in large $\mu$ cases. Here, $\tan\beta=8$ and $\delta m_{H_u}^2=\delta m_{H_d}^2=0$.
In the orange (yellow) region, the muon $g-2$ is explained at 1$\sigma$ (2$\sigma$) level. It is denoted that $m_{L}(M_{\rm GUT}) = {\rm sign}(m_L^2) \sqrt{|m_L^2|}|_{\rm M_{\rm GUT}}$.
We take $m_t({\rm pole})=173.34$ GeV and $\alpha_s(m_Z)=0.1185$.
}
\label{fig:largemu}
\end{center}
\end{figure}
Here, we consider the model with non-zero $M_{1/2}$. 
In this case, there is a region where $(\delta a_\mu)_{\tilde{B}-{\tilde \mu}_L-{\tilde \mu}_R}$ dominates $(\delta a_\mu)_{\rm SUSY}$. To obtain $(\delta a_\mu)_{\tilde{B}-{\tilde \mu}_L-{\tilde \mu}_R} \gtrsim 1.8 \cdot 10^{-9}$, it is required that $\mu$ is as large as $\sim 3$ TeV and the smuons and bino are as light as $100$\,-\,$300$ GeV.

In large $\mu$ case, the Higgs soft masses are not required to be tuned for realizing successful electroweak symmetry breaking; therefore, we set $\delta m_{H_u}^2=\delta m_{H_d}^2=0$, for simplicity. In Fig.~\ref{fig:largemu}, we show the contours of the Higgs boson mass and the region explaining $\Delta a_\mu$. Here, $\tan\beta=8$. %
We take $M_1(M_{\rm GUT})$ as an input parameter instead of $\delta M_{1/2}$. 
Also, $m_{\bar E}(M_{\rm GUT})$ and $m_{\bar L}(M_{\rm GUT})$ are input parameters, which corresponds to choosing $m_{\bar 5}^2$ and $m_{10}^2$. The sign of $\mu$ is chosen such that $(\delta a_\mu)_{\rm SUSY}$ is positive (same sign of the bino mass). One can see that there is a region where the discrepancy of the muon $g-2$ from the SM prediction is reduced to 1$\sigma$ level (orange) for $m_{L}^2(M_{\rm GUT})<0$. 
The negative soft mass squared at the GUT scale is required, since the wino mass is rather large and it gives large positive radiative correction to the left-handed slepton masses: to make the left-handed sleptons light, the fine-tuning of $m_{L}^2(M_{\rm GUT})$ is needed.
Consequently, in the case $M_1<0$ (right panel), the region which can explain $\Delta a_\mu$ is smaller due to the larger wino mass, compared to the case $M_1>0$ (left panel). 
The gray region is excluded since the stau becomes LSP. On the edge of the gray region, the relic abundance of the lightest neutralino explains the observed value of the dark matter, $\Omega_{\rm DM} h^2 \simeq 0.12$~\cite{relic_obs}, via the coannihilation with the stau~\cite{coan}.

Also, we show sample mass spectra of two model points in Table \ref{table:spectrum2}. The squark and gluino are heavier than the previous case, 
$\delta M_{1/2}=0$: the masses of the squarks and gluino are 3\,-\,4.5\,TeV. However, it may be still possible to discover or exclude them at the LHC with an integrated luminosity of $3000$\,fb$^{-1}$. On the other hand, the direct production of the sleptons are more promising to be checked, since they can not be much heavier than 300 GeV for explaining $\Delta a_\mu$.

\begin{table}[t!]
    \caption{The mass spectra for large $\mu$ case. Here,  $\delta M_{1/2} \neq 0$, $\delta m_{H_u}^2= \delta m_{H_d}^2=0$.
   }
  \begin{center}
    \begin{tabular}{  c | c  }
            {\bf P3} & \\
\hline
    $m_{3/2}$ & 70 TeV \\
    $M_1(M_{\rm GUT})$ & 230 GeV \\
    $m_{\bar E}(M_{\rm GUT})$ & 230 GeV \\
    $m_{L}(M_{\rm GUT})$ & -550 GeV \\
    $\delta A_t(M_{\rm GUT})$ & 2600 GeV\\
    $\tan \beta$ & 8 \\
    \hline
\hline    
    $m_{\rm gluino}$ & 3.8 TeV \\
      $m_{\tilde{q}}$ & 3.5 TeV \\
      $\mu$ & 3.2 TeV\\
    $m_{\tilde{t}_{1,2}}$ & 2.5, 3.3 TeV \\
    $m_{\tilde{e}_L} (m_{\tilde{\mu}_L})$ & 310 GeV\\
    $m_{\tilde{e}_R} (m_{\tilde{\mu}_R})$ & 236 GeV\\
    $m_{\tilde{\tau}_1}$ & 130 GeV\\
     $m_{\chi_1^0}$, $m_{\chi_2^0}$ & 112, 817 GeV \\
     $m_{\chi_3^0}$, $m_{\chi_4^0}$ & 3197, 3197 GeV \\
     $m_{\chi_1^{\pm}}$, $m_{\chi_2^{\pm}}$ & 817, 3197 GeV \\
     $m_{h}$ & 123.9 GeV \\
    $ (\delta a_\mu)_{\rm SUSY}$ & $1.80 \cdot 10^{-9}$ \\
    \end{tabular}
        \hspace{20pt}
        \begin{tabular}{  c | c  }
                    {\bf P4} & \\
\hline
    $m_{3/2}$ & 70 TeV \\
    $M_1(M_{\rm GUT})$ & -300 GeV \\
    $m_{\bar E}(M_{\rm GUT})$ & 265 GeV \\
    $m_{L}(M_{\rm GUT})$ & -920 GeV \\
    $\delta A_t(M_{\rm GUT})$ & 2200 GeV\\
    $\tan \beta$ & 8 \\
    \hline
\hline    
    $m_{\rm gluino}$ & 4.8 TeV \\
      $m_{\tilde{q}}$ & 4.3 TeV \\
      $\mu$ & -3.6 TeV \\
    $m_{\tilde{t}_{1,2}}$ & 3.3, 4.2 TeV \\
    $m_{\tilde{e}_L} (m_{\tilde{\mu}_L})$ &314 GeV\\
    $m_{\tilde{e}_R} (m_{\tilde{\mu}_R})$ & 273 GeV\\
    $m_{\tilde{\tau}_1}$ & 123 GeV\\
     $m_{\chi_1^0}$, $m_{\chi_2^0}$ & 110, 1242 GeV \\
     $m_{\chi_3^0}$, $m_{\chi_4^0}$ & 3570, 3571 GeV \\
     $m_{\chi_1^{\pm}}$, $m_{\chi_2^{\pm}}$ & 1242, 3571 GeV \\
     $m_{h}$ & 125.1 GeV \\
    $ (\delta a_\mu)_{\rm SUSY}$ & $1.84 \cdot 10^{-9}$ \\
    \end{tabular}
  \label{table:spectrum2}
  \end{center}
\end{table}

\section{A realization of the pAMSB} \label{sec:pamsbr}

We consider a more fundamental realization of the pAMSB, motivated by the mixed modulus-anomaly mediation scenario~\cite{mirage1, endo}.
Here, we consider the following K\"{a}hler potential and superpotential: 
\begin{eqnarray}
K &=& -3 \ln (-f/3), \nonumber \\
f &\ni& (X+X^\dag)^{n_{10}}(Q^\dag Q + \bar{U}^\dag \bar{U} + \bar{E}^\dag \bar{E}) \nonumber \\
&+& (X+X^\dag)^{n_{5}}(L^\dag L + \bar{D}^\dag \bar{D})  \nonumber \\
&+& (X+X^\dag)^{n_{u}}(H_u^\dag H_u) + (X+X^\dag)^{n_{d}}(H_d^\dag H_d), \nonumber \\
W &=& -A e^{-b X} + w(Z),
\end{eqnarray}
where we have taken the unit of $M_P=1$ and the MSSM matter superfields couple to a  moduli field $X$ in the K\"{a}hler potential.
The superpotential for a SUSY breaking field $Z$, $w(Z)$, contains a constant term, which is around the gravitino mass $m_{3/2}$. 
The moduli $X$ has a $F$-term of $\left<F_X\right>/(2\, {\rm Re}\left<X\right>) \sim m_{3/2}/100$: corrections to the soft SUSY breaking masses are comparable with those from anomaly mediation. The SUSY breaking de Sitter vacuum is obtained thanks to a coupling between $X$ and $Z$, $f \ni (X+X^\dag)^{s+1} |Z|^2$~\cite{endo}.
The detailed explanations are shown in Appendix \ref{sec:mirage-like}.
It is also assumed that $X$ couples to the field strength superfield of the vector multiplets, giving tree level gaugino masses.
Then, together with contributions from anomaly mediation, the soft SUSY breaking parameters are obtained as
\begin{eqnarray}
m_{k}^2 &=& n_k \left|\frac{\left<F_X\right>}{\left<x\right>}\right|^2 + (m_{k}^2)_{\rm AMSB} + (m_{k}^2)_{\rm mixed}, \nonumber \\
M_{a} &=& \delta M_{1/2} + \frac{\beta_a}{g_a} m_{3/2}, \nonumber \\
A_t &=& -(n_{10} + n_{10} + n_{u}) \frac{\left<F_X\right>}{\left<x\right>} - (\beta_{Y_t}/Y_t) {m_{3/2}}, \nonumber \\
A_b &=& -(n_{10} + n_{5} + n_{d}) \frac{\left<F_X\right>}{\left<x\right>} - (\beta_{Y_b}/Y_b) {m_{3/2}}, \nonumber \\
A_{\tau} &=& -(n_{10} + n_{5} + n_{d}) \frac{\left<F_X\right>}{\left<x\right>} - (\beta_{Y_{\tau}}/Y_{\tau}) {m_{3/2}}, 
\end{eqnarray}
where  $(m_{i}^2)_{\rm AMSB}$ is a contribution from anomaly mediation and $(m_{i}^2)_{\rm mixed}$ is a mixed  contribution from the moduli and anomaly mediation. Here, $x = X + X^\dag$.
The detailed mass formulae are shown in Eq.(\ref{eq:mirage1}) in Appendix \ref{sec:mirage-like}.
In this model, we can write the soft SUSY breaking masses using the following parameters:
\begin{eqnarray}
\left[n_{10}, \ n_{5},  \ n_u,  \ n_d,  \ \delta M_{1/2},  \ {m_{3/2}},  \ \frac{\left<F_X\right>}{\left<x\right>} \right]. \label{eq:mirage_para}
\end{eqnarray}
With these parameters, we can easily reproduce the results of the large $\mu$ case. 

However, it is difficult to accommodate the small $\mu$ cases.
When $\mu$ is as small as $\sim 100$\,GeV, the large contribution to $m_{H_u}^2$ from the moduli, $n_u |\left<F_X\right> /\left<x\right>|^2$, is required: the Higgs potential has to be tuned with $m_{H_u}^2$ rather than $\mu^2$ such that the observed electroweak symmetry breaking (EWSB) scale is generated. 
As a result, the trilinear coupling $A_{u,c,t} \sim n_u \left<F_X\right> /\left<x\right>$ becomes large, and a color breaking vacuum deeper than the EWSB minimum may be generated~\cite{ccb} (see also \cite{ccb_recent} for a recent discussion).\footnote{
Roughly, to avoid the constraint from the color breaking minimum, the condition
\begin{eqnarray}
 3 (m_{Q}^2 + m_{\bar{U}}^2)  \sim 6\, n_u |\left<F_X\right> /\left<x\right>|^2 > n_u^2 |\left<F_X\right> /\left<x\right>|^2   \Rightarrow  n_u < 6, \notag
\end{eqnarray}
should be satisfied. Here, we estimate the radiative correction to the Higgs soft mass squared, $\Delta m_{H_u}^2$, as $\sim (3Y_t^2/4\pi^2)\, m_{Q}^2\, \ln (M_{\rm GUT}/m_{\rm SUSY})$ and require that $\Delta m_{H_u}^2$ be canceled by $n_u |\left<F_X\right> /\left<x\right>|^2$. 
In this case, the small $\mu$ is realized only when $m_0$ is fairly large. Therefore, it is difficult to explain the muon $g-2$ anomaly unless $|n_5| \ll 1$: the left-handed slepton is not light enough anymore.
}

Moreover, this large $A$-term, $A_t \sim -n_u \left<F_X\right>/\left<x\right>$, does not help to enhance the Higgs boson mass: if $A_t$ is positive, the stop tends to be tachyonic due to $(m^2_k)_{\rm mixed}$. On the other hand, if $A_t$ is negative, it is destructive to the radiative correction from the gluino and the weak scale value of $A_t$ is not large anymore.

\vspace{10pt}
To accommodate the small $\mu$ case, i.e. generating large $m_{H_u}^2$ without inducing too large $A_{u,c,t}$, we consider the following interaction:
\begin{eqnarray}
W &=& \lambda_Y Y_1 H_u H_d + M_{Y} Y_1 Y_2 + \frac{\kappa}{2} Z Y_1^2,
\end{eqnarray}
%
%
where $Y_1$ and $Y_2$ are heavy fields, and $\kappa \left<Z\right> \ll M_Y$ is assumed. We take the K\"{a}hler potential for $Y_1$ and $Y_2$ as $K\ni Y_1^\dag Y_1 + Y_2^\dag Y_2 + ({\rm higher\,\, powers\,\, of \,\,} Y_1^\dag Y_1{\rm \,\, and\,\,} Y_2^\dag Y_2)$.
The above interaction is consistent with the $R$-symmetry, where the $R$-charges are assigned as $R(H_u H_d)=R(Y_2)=R(Z)=2$ and $R(Y_1)=0$. Then, tree level gaugino masses from $Z$ are  prohibited.\footnote{
The shift-symmetry breaking term in the superpotential, $W \ni A e^{-bX}$, is consistent with the $R$-symmetry, 
if $X$ transforms as $X \to X - 2i\theta_R/b$. In this case, the moduli contribution to the gaugino masses is also prohibited, which corresponds to $\delta M_{1/2}=0$. However, as shown in Sec.\,\ref{seq:small_mu}, the muon $g-2$ anomaly can be successfully explained with $\delta M_{1/2}=0$ in the small $\mu$ case.
}

After integrating out  $Y_1$ and $Y_2$, the one-loop soft masses for the Higgs doublets are generated as 
\begin{eqnarray}
\delta' m_{H_u}^2 = \delta' m_{H_d}^2 \simeq \frac{\lambda_Y^2}{32\pi^2}\frac{|\kappa F_Z|^2}{M_{Y}^2}, \label{eq:additional2}
\end{eqnarray}
at the leading order. For instance, taking $M_{Y}=10^{15}$\,GeV,\footnote{
Here, we recover the unit of $M_P \simeq 2.4 \cdot 10^{18}$\,GeV.
}
 $\kappa=0.08$, and $\lambda_Y=10^{-3}$, we have desired size of $\delta' m^2_{H_{u,d}} \simeq 10^{-4}\, m_{3/2}^2$. On the other hand, the generated $A$-terms and the Higgs $B$-term are $\sim \lambda_Y^2/(16\pi^2)\, m_{3/2}$, which pick up  $B_Y$ ($B_Y$ is the $B$-term of $Y_1 Y_2$); therefore they are suppressed compared to $(\delta' m^2_{H_{u,d}})^{1/2}$.
Including $\delta' m_{H_u}^2$ and $\delta' m_{H_d}^2$ in Eq.(\ref{eq:additional2}), together with the parameters in Eq.(\ref{eq:mirage_para}), we can reproduce the SUSY mass spectrum of the pAMSB almost completely.

\section{Conclusion and discussion} \label{sec:conc}

We have proposed a simple anomaly mediation model, namely the phenomenological anomaly mediated SUSY breaking  (pAMSB) model, in order to explain the Higgs boson mass around 125 GeV and the muon $g-2$ anomaly. The pAMSB can be regarded as a generalization of mixed modulus-anomaly mediation. 
We have shown that the muon $g-2$ anomaly and the observed Higgs boson mass are easily explained. Moreover, our model can be accommodated into $SU(5)$ or $SO(10)$ GUT without difficulty, since required GUT breaking effects to  obtain the mass splitting among the strongly and weakly interacting SUSY particles are induced by anomaly mediation. We have also presented a possible realization of the pAMSB. 

When the muon $g-2$ anomaly is explained by the wino-Higgsino-(muon sneutrino) diagram, the gluino and squark masses can be as small as $2$\,-\,$3$ TeV; therefore our scenario is expected to be tested at the LHC with $\sqrt{s}=14$ TeV. Even in the other case, where the $\tilde{B}-{\tilde \mu}_L-{\tilde \mu}_R$ diagram dominates the SUSY contribution, the sleptons masses are around 300 GeV, and hence, the existence of the these light sleptons can be checked easily. 

Finally let us briefly comment on the cosmological aspects of the pAMSB. 
Since the gravitino is as heavy as $\sim 100$ TeV, the cosmological gravitino problem is relaxed.
In our model, there exists the moduli field $X$, which lifts up the slepton masses via its $F$-term. 
The decay of the moduli into the gravitinos with a large branching fraction may spoil the success of the standard cosmology and may be problematic~\cite{moduli1}; however, it can be solved if the moduli strongly couples to the inflaton~\cite{linde, ty_nty}.

\section*{Acknowledgments}
We thank Luca Silvestrini for useful discussion and careful reading of the manuscript. The research leading to these results has received funding from the European Research Council under the European Unions Seventh Framework Programme (FP/2007-2013)/ERC Grant Agreement N\textsuperscript{o} 279972
``NPFlavour''. 

\appendix
\section{Soft mass parameters} \label{sec:appsoftmass}
In this appendix, we list the formulae for the soft mass parameters. We use the unit where the reduced Planck mass is set to unity in the following discussions.

\subsection{AMSB} \label{eq:amsb}
The soft SUSY breaking parameters with a sequestered K\"{a}hler potential are listed. Here, we consider the case that there is no tree level gaugino mass term.
The scalar masses from anomaly mediation are~\cite{amsb2}
\beq
{m}_{Q_i}^{\prime \, 2} &=&  \left[-\frac{8}{3} g_3^4 b_3 - \frac{3}{2}g_2^4 b_2  - \frac{1}{30} g_1^4 b_1 + \delta_{i 3} (16\pi^2) (Y_t \beta_{Y_t} + Y_b \beta_{Y_b}) \right] \frac{m_{3/2}^2}{(16\pi^2)^2}, \nonumber \\
m_{\bar{U}_i}^{\prime \,2} &=&  \left[-\frac{8}{3} g_3^4 b_3  - \frac{8}{15}g_1^4 b_1 + \delta_{i 3} (16\pi^2) 2 Y_t \beta_{Y_t} \right] \frac{m_{3/2}^2}{(16\pi^2)^2},  \nonumber \\
m_{{\bar{D}}_i}^{\prime \,2} &=& \left[-\frac{8}{3} g_3^4 b_3  - \frac{2}{15}g_1^4 b_1 + \delta_{i 3} (16\pi^2) 2Y_b \beta_{Y_b} \right] \frac{m_{3/2}^2}{(16\pi^2)^2},  \nonumber \\
m_{{{L}}_i}^{\prime \,2} &=&  \left[- \frac{3}{2}g_2^4 b_2  - \frac{3}{10} g_1^4 b_1 + \delta_{i 3} (16\pi^2) Y_{\tau} \beta_{Y_\tau} \right] \frac{m_{3/2}^2}{(16\pi^2)^2},\nonumber \\
m_{{\bar{E}}_i}^{\prime \,2} &=& \left[ -\frac{6}{5} g_1^4 b_1 + \delta_{i 3} (16\pi^2)  2 Y_{\tau} \beta_{Y_\tau} \right] \frac{m_{3/2}^2}{(16\pi^2)^2}, \nonumber \\
m_{{H_u}}^{\prime \,2} &=&   \left[- \frac{3}{2}g_2^4 b_2  - \frac{3}{10}g_1^4 b_1 + (16\pi^2) 3 Y_{t} \beta_{Y_t} \right] \frac{m_{3/2}^2}{(16\pi^2)^2},  \nonumber \\
m_{{H_d}}^{\prime \,2} &=&   \left[- \frac{3}{2}g_2^4 b_2  - \frac{3}{10}g_1^4 b_1 + (16\pi^2) (Y_{\tau} \beta_{Y_\tau} + 3 Y_{b} \beta_{Y_b}) \right] \frac{m_{3/2}^2}{(16\pi^2)^2}, \label{eq:mass_amsb}
\eeq
where $b_i$ are the coefficients of the one-loop beta-functions for gauge couplings: $b_i=(33/5, 1, -3)$. For third generation sfermions, there are terms proportional to the Yukawa couplings and their beta-function. Here, we have neglected first and second generation Yukawa couplings.
The gaugino masses are given by
\beq
M_{1} = \frac{33}{5} g_1^2 \frac{m_{3/2}}{16\pi^2}, \ M_{2} = g_2^2 \frac{m_{3/2}}{16\pi^2}, \ M_{3} = -3 g_3^2 \frac{m_{3/2}}{16\pi^2} ,
\eeq
at the one-loop level. Trilinear couplings are given by
\beq
A_{t} = -(\beta_{Y_t}/Y_t) {m_{3/2}}, \ \ A_{b} = -(\beta_{Y_b}/Y_b) {m_{3/2}}, \ \ A_{\tau} = -(\beta_{Y_{\tau}}/Y_{\tau}) {m_{3/2}}.
\eeq

\subsection{A model with KKLT type potential} \label{sec:mirage-like}
Following Ref.~\cite{endo}, we consider the following K\"{a}hler potential and superpotential:
\begin{eqnarray}
K &=& -3 \ln (-f/3), \nonumber \\
f &=& -3 (X + X^\dag) + c_Z (X+X^\dag)^{s+1} |Z|^2, \nonumber \\
W &=& -Ae^{-bX} + w(Z),
\end{eqnarray}
where $X$ is a moduli field and $Z$ is a SUSY breaking field.
 The superpotential for $Z$ is denoted by $w(Z)$, which contains the constant term: $w(Z=0)=\mathcal{C}$. 
The parameter $A$ and constant term $\mathcal{C}$ are taken to be real positive by the shift of $X$ and $U(1)_R$ transformation without loss of generality. 

Provided $\left<Z\right> \ll 1$,\footnote{
Unlike the Polonyi field, the SUSY breaking field $Z$ is not necessarily a gauge singlet: the origin may be ensured by a symmetry.
}
the relevant part of the K\"{a}hler potential is written as
\begin{eqnarray}
K = -3 \ln x+c_Z x^s |Z|^2 + \dots ,
\end{eqnarray}
where $x=X+X^\dag$. Then, the scalar potential is given by
\begin{eqnarray}
V = \frac{A b e^{-bx}}{3x^2} \Bigl[Abx + 6A - 6\mathcal{C}e^{bx/2} \cos(b\, {\rm Im}(X)) \Bigr] + \left|\frac{\partial w}{\partial Z}\right|^2\frac{x^{-s-3}}{c_Z}
\end{eqnarray}
The imaginary part of $X$ is stabilized at ${\rm Im}(X)=0$, and the scalar potential for $x$ is
\begin{eqnarray}
V = \frac{A b e^{-bx}}{3x^2} \Bigl[Abx + 6A - 6\mathcal{C}e^{bx/2}  \Bigr] + \frac{D}{x^{s'}},
\end{eqnarray}
where $s'=s+3$ and $D=|\partial w(Z)/\partial Z|^2$. 
Using the minimization condition $(\partial V /\partial x)=0$ and the condition for the vanishing cosmological constant $V=0$, the minimum is found for $b \left<x\right> \sim 70$ with the equation:
\begin{eqnarray}
3 \mathcal{C} e^{y/2} (4- 2 s' +y)+A[-12-7y-y^2+s'(6+y)]=0,
\end{eqnarray}
where $y=bx$. Here, we consider the case of $\mathcal{C}\sim10^{-13}$ and $A \sim 1$.
We see that $\left<F_X\right>/\left<x\right>$ is suppressed by a factor $y \sim 70$ compared to the gravitino mass.
\begin{eqnarray}
\frac{\left<F_X\right>}{\left<x\right>} \simeq e^{K/2} \mathcal{C} \left[\frac{y s'}{(y+3)(y+4) -s' (6+y)}\right] \sim  \frac{s' m_{3/2}}{70}.
\end{eqnarray}
Note that further suppression is possible if one consider more general K\"{a}hler potential and super potential for $X$~\cite{kklt_general}.

\vspace{20pt}
Now, we couple $X$ to the matter fields such that the soft SUSY breaking masses which are comparable to those from anomaly mediation are obtained. The couplings are given by
\begin{eqnarray}
\Delta f &=& (X+X^\dag)^{n_{10}}(Q^\dag Q + \bar{U}^\dag \bar{U} + \bar{E}^\dag \bar{E}) \nonumber \\
&+& (X+X^\dag)^{n_{5}}(L^\dag L + \bar{D}^\dag \bar{D})  \nonumber \\
&+& (X+X^\dag)^{n_{u}}(H_u^\dag H_u) + (X+X^\dag)^{n_{d}}(H_d^\dag H_d).
\end{eqnarray}
The K\"{a}hler potential is replaced as $K=-3 \ln [-(f+\Delta f)/3]$. 
The canonically normalized $Q_k$ is obtained by $Q_k^c = [\left<x\right>^{n_k-1}]^{1/2} Q_k$.
Then, scalar masses at the tree level are
\begin{eqnarray}
m_{Q}^2 = m_{\bar U}^2 = m_{\bar E}^2 = n_{10} \frac{|\left<F_X\right>|^2}{\left<x\right>^2}, \\
m_{L}^2 = m_{\bar D}^2 = n_{5} \frac{|\left<F_X\right>|^2}{\left<x\right>^2}, \\
m_{H_u}^2 = n_{u} \frac{|\left<F_X\right>|^2}{\left<x\right>^2},  \ m_{H_d}^2 = n_{d} \frac{|\left<F_X\right>|^2}{\left<x\right>^2}. 
\end{eqnarray}
The trilinear couplings are given by
\begin{eqnarray}
A_u = (n_{10} + n_{10} + n_{u}) \frac{\left<F_X\right>}{\left<x\right>},  \ A_d =A_e =  (n_{10} + n_{5} + n_{d}) \frac{\left<F_X\right>}{\left<x\right>}.
\end{eqnarray}
The gaugino masses are generated from the gauge kinetic functions:
\begin{eqnarray}
\int d^2\theta \frac{1}{4} X^l W_{\alpha} W^{\alpha} + h.c. =\int d^2\theta \frac{1}{4} \left<X\right>^l\left( 1 + l \frac{\left<F_X\right>}{\left<X\right>} \theta^2\right) W_{\alpha} W^{\alpha}  + h.c.\, ,
\end{eqnarray}
and
\begin{eqnarray}
M_\lambda = -\frac{l}{2} \frac{\left<F_X\right>}{\left<X\right>}.
\end{eqnarray}
Here, ${\rm Re}\left<X\right>^l = 1/g^2$.
Including the contributions from AMSB, we obtain
\begin{eqnarray}
m_{k}^2 &=& n_k \left|\frac{\left<F_X\right>}{\left<x\right>}\right|^2 + (m_{k}^2)_{\rm AMSB} + (m_{k}^2)_{\rm mixed}, \nonumber \\
M_{a} &=& \frac{l}{2} \frac{\left<F_X\right>}{\left<X\right>} + \frac{\beta_a}{g_a} m_{3/2}  = \delta M_{1/2} + \frac{\beta_a}{g_a} m_{3/2} \nonumber \\
A_t &=& -(n_{10} + n_{10} + n_{u}) \frac{\left<F_X\right>}{\left<x\right>} - (\beta_{Y_t}/Y_t) {m_{3/2}}, \nonumber \\
A_b &=& -(n_{10} + n_{5} + n_{d}) \frac{\left<F_X\right>}{\left<x\right>} - (\beta_{Y_b}/Y_b) {m_{3/2}}, \nonumber \\
A_{\tau} &=& -(n_{10} + n_{5} + n_{d}) \frac{\left<F_X\right>}{\left<x\right>} - (\beta_{Y_{\tau}}/Y_{\tau}) {m_{3/2}}, \label{eq:mirage1}
\end{eqnarray}
where $(m_{k}^2)_{\rm AMSB}$ is the contribution coming purely from AMSB shown in Eq.(\ref{eq:mass_amsb}), and $(m_k^2)_{\rm mixed}$ is
\begin{eqnarray}
(m_k^2)_{\rm mixed} &=& \frac{1}{2} \, \frac{m_{3/2}}{16\pi^2} \Bigl[
c_a^k g_a^2 \left(-\delta M_{1/2} + h.c.\right) \nonumber \\
&+& \sum_{lm} ((n_k + n_l + n_m) \frac{\left<F_X\right>}{\left<x\right>} + h.c.) d^{\, k} |y_{klm}|^2 \label{eq:mirage2}
\Bigr]. 
\end{eqnarray}
Here, we have flipped the signs of $A$-terms and $M_i$ by the $U(1)_R$ rotation.
The coefficients $c_a^k$ and $d^k$ can be read from the anomalous dimension of the field $k$:
\begin{eqnarray}
\gamma_k \equiv \frac{\partial \ln Z_k}{\partial \ln \mu} = \frac{1}{16\pi^2} (c_a^k g_a^2 - d^{\, k} \sum_{lm} |y_{klm}|^2).
\end{eqnarray}


\begin{thebibliography}{99}

\bibitem{gm2_exp}
  G.~W.~Bennett {\it et al.}  [Muon g-2 Collaboration],
  Phys.\ Rev.\ D {\bf 73}, 072003 (2006)
  [hep-ex/0602035];
 B.~L.~Roberts,
  Chin.\ Phys.\ C {\bf 34}, 741 (2010)
  [arXiv:1001.2898 [hep-ex]].

\bibitem{gm2_hagiwara}
  K.~Hagiwara, R.~Liao, A.~D.~Martin, D.~Nomura and T.~Teubner,
  J.\ Phys.\ G {\bf 38}, 085003 (2011)
  [arXiv:1105.3149 [hep-ph]].
  
\bibitem{gm2_davier}
 M.~Davier, A.~Hoecker, B.~Malaescu and Z.~Zhang,
  Eur.\ Phys.\ J.\ C {\bf 71}, 1515 (2011)
  [Erratum-ibid.\ C {\bf 72}, 1874 (2012)]
  [arXiv:1010.4180 [hep-ph]].
  
\bibitem{gm2_sm_ew}
  A.~Czarnecki, W.~J.~Marciano and A.~Vainshtein,
  Phys.\ Rev.\ D {\bf 67}, 073006 (2003)
  [Erratum-ibid.\ D {\bf 73}, 119901 (2006)]
  [hep-ph/0212229].
  
\bibitem{gm2_susy1}
 J.~L.~Lopez, D.~V.~Nanopoulos and X.~Wang,
  Phys.\ Rev.\ D {\bf 49}, 366 (1994)
  [hep-ph/9308336];
U.~Chattopadhyay and P.~Nath,
  Phys.\ Rev.\ D {\bf 53}, 1648 (1996)
  [hep-ph/9507386].

\bibitem{gm2_susy2} 
  T.~Moroi,
  Phys.\ Rev.\ D {\bf 53}, 6565 (1996)
  [Erratum-ibid.\ D {\bf 56}, 4424 (1997)]
  [hep-ph/9512396].
  
\bibitem{lhc_susy}
  S.~Chatrchyan {\it et al.}  [CMS Collaboration],
  JHEP {\bf 1406}, 055 (2014)
  [arXiv:1402.4770 [hep-ex]];
  G.~Aad {\it et al.}  [ATLAS Collaboration],
  JHEP {\bf 1409}, 176 (2014)
  [arXiv:1405.7875 [hep-ex]].


\bibitem{lhc_higgs}
 G.~Aad {\it et al.}  [ATLAS and CMS Collaborations],
  arXiv:1503.07589 [hep-ex].
  

\bibitem{higgs_susy}
  Y.~Okada, M.~Yamaguchi, T.~Yanagida,
  Prog.\ Theor.\ Phys.\  {\bf 85 } (1991)  1-6;
  J.~R.~Ellis, G.~Ridolfi, F.~Zwirner,
  Phys.\ Lett.\  {\bf B257 } (1991)  83-91;
  H.~E.~Haber, R.~Hempfling,
  Phys.\ Rev.\ Lett.\  {\bf 66 } (1991)  1815-1818.
  
 \bibitem{higgs_3loop}
   T.~Hahn, S.~Heinemeyer, W.~Hollik, H.~Rzehak and G.~Weiglein,
  Phys.\ Rev.\ Lett.\  {\bf 112}, no. 14, 141801 (2014)
  [arXiv:1312.4937 [hep-ph]].


\bibitem{split_gm2_1}
  M.~Ibe, T.~T.~Yanagida and N.~Yokozaki,
  JHEP {\bf 1308}, 067 (2013)
  [arXiv:1303.6995 [hep-ph]].

\bibitem{split_gm2_2}
   K.~S.~Babu, I.~Gogoladze, Q.~Shafi and C.~S.~Ün,
  Phys.\ Rev.\ D {\bf 90}, no. 11, 116002 (2014)
  [arXiv:1406.6965 [hep-ph]].
  
  \bibitem{vector}
  M.~Endo, K.~Hamaguchi, S.~Iwamoto and N.~Yokozaki,
  Phys.\ Rev.\ D {\bf 84} (2011) 075017
  [arXiv:1108.3071 [hep-ph]];
   M.~Endo, K.~Hamaguchi, S.~Iwamoto and N.~Yokozaki,
  Phys.\ Rev.\ D {\bf 85} (2012) 095012
  [arXiv:1112.5653 [hep-ph]];
  M.~Endo, K.~Hamaguchi, K.~Ishikawa, S.~Iwamoto and N.~Yokozaki,
  JHEP {\bf 1301}, 181 (2013)
  [arXiv:1212.3935 [hep-ph]].
  
  \bibitem{gm2_gauge}
    R.~Sato, K.~Tobioka and N.~Yokozaki,
  Phys.\ Lett.\ B {\bf 716}, 441 (2012)
  [arXiv:1208.2630 [hep-ph]];
    M.~Ibe, S.~Matsumoto, T.~T.~Yanagida and N.~Yokozaki,
  JHEP {\bf 1303}, 078 (2013)
  [arXiv:1210.3122 [hep-ph]];
   G.~Bhattacharyya, B.~Bhattacherjee, T.~T.~Yanagida and N.~Yokozaki,
  Phys.\ Lett.\ B {\bf 725}, 339 (2013)
  [arXiv:1304.2508 [hep-ph]];
   G.~Bhattacharyya, B.~Bhattacherjee, T.~T.~Yanagida and N.~Yokozaki,
  Phys.\ Lett.\ B {\bf 730} (2014) 231
  [arXiv:1311.1906 [hep-ph]].
  
 \bibitem{gm2_gaugino}
 S.~Iwamoto, T.~T.~Yanagida and N.~Yokozaki,
  arXiv:1407.4226 [hep-ph];
  K.~Harigaya, T.~T.~Yanagida and N.~Yokozaki,
  Phys.\ Rev.\ D {\bf 91}, no. 7, 075010 (2015)
  [arXiv:1501.07447 [hep-ph]].

 \bibitem{gm2_gaugino2}
  K.~Harigaya, T.~T.~Yanagida and N.~Yokozaki,
  arXiv:1505.01987 [hep-ph].



  
\bibitem{DT_split}
  T.~Yanagida,
  Phys.\ Lett.\ B {\bf 344}, 211 (1995)
  [hep-ph/9409329];
  T.~Hotta, K.~I.~Izawa and T.~Yanagida,
  Phys.\ Rev.\ D {\bf 53}, 3913 (1996)
  [hep-ph/9509201].
\bibitem{DT_split2}
  E.~Witten,
  hep-ph/0201018.


  
  \bibitem{gm2_gravity}
   S.~Mohanty, S.~Rao and D.~P.~Roy,
  JHEP {\bf 1309}, 027 (2013)
  [arXiv:1303.5830 [hep-ph]];
    S.~Akula and P.~Nath,
  Phys.\ Rev.\ D {\bf 87}, no. 11, 115022 (2013)
  [arXiv:1304.5526 [hep-ph]];
   J.~Chakrabortty, S.~Mohanty and S.~Rao,
  JHEP {\bf 1402}, 074 (2014)
  [arXiv:1310.3620 [hep-ph]];
    I.~Gogoladze, F.~Nasir, Q.~Shafi and C.~S.~Un,
  Phys.\ Rev.\ D {\bf 90}, no. 3, 035008 (2014)
  [arXiv:1403.2337 [hep-ph]];
    M.~A.~Ajaib, I.~Gogoladze and Q.~Shafi,
  arXiv:1501.04125 [hep-ph].

  
\bibitem{amsb1}
   G.~F.~Giudice, M.~A.~Luty, H.~Murayama and R.~Rattazzi,
  JHEP {\bf 9812}, 027 (1998)
  [hep-ph/9810442].
\bibitem{amsb2}
 L.~Randall and R.~Sundrum,
  Nucl.\ Phys.\ B {\bf 557}, 79 (1999)
  [hep-th/9810155].
  
 \bibitem{note_added} 
  F.~Wang, W.~Wang, J.~M.~Yang and Y.~Zhang,
  arXiv:1505.02785 [hep-ph].
  
  
\bibitem{kklt}
S.~Kachru, R.~Kallosh, A.~D.~Linde and S.~P.~Trivedi,
  Phys.\ Rev.\ D {\bf 68}, 046005 (2003)
  [hep-th/0301240].
  
\bibitem{mamsb}
  T.~Gherghetta, G.~F.~Giudice and J.~D.~Wells,
  Nucl.\ Phys.\ B {\bf 559}, 27 (1999)
  [hep-ph/9904378].
  
  \bibitem{mamsb_2} 
 J.~L.~Feng and T.~Moroi,
  Phys.\ Rev.\ D {\bf 61}, 095004 (2000)
  [hep-ph/9907319];
  U.~Chattopadhyay, D.~K.~Ghosh and S.~Roy,
  Phys.\ Rev.\ D {\bf 62}, 115001 (2000)
  [hep-ph/0006049].
  
  
  \bibitem{amsb_other} 
  A.~Arbey, A.~Deandrea, F.~Mahmoudi and A.~Tarhini,
  Phys.\ Rev.\ D {\bf 87}, no. 11, 115020 (2013)
  [arXiv:1304.0381 [hep-ph]].



\bibitem{photonic}
  G.~Degrassi and G.~F.~Giudice,
  Phys.\ Rev.\ D {\bf 58}, 053007 (1998)
  [hep-ph/9803384].

\bibitem{photonic_recent} 
  P.~von Weitershausen, M.~Schafer, H.~Stockinger-Kim and D.~Stockinger,
  Phys.\ Rev.\ D {\bf 81}, 093004 (2010)
  [arXiv:1003.5820 [hep-ph]].


\bibitem{suspect}
    A.~Djouadi, J.~-L.~Kneur and G.~Moultaka,
  Comput.\ Phys.\ Commun.\  {\bf 176}, 426 (2007)
  [hep-ph/0211331].

\bibitem{feynhiggs}
    S.~Heinemeyer, W.~Hollik and G.~Weiglein,
  Comput.\ Phys.\ Commun.\ \ {\bf 124}, 76  (2000)
  [hep-ph/9812320];
  Eur.\ Phys.\ J.\ C\ {\bf 9}, 343  (1999)
  [hep-ph/9812472];
  G.~Degrassi, S.~Heinemeyer, W.~Hollik, P.~Slavich and G.~Weiglein,
  Eur.\ Phys.\ J.\ C\ {\bf 28}, 133  (2003)
  [hep-ph/0212020];
  M.~Frank, T.~Hahn, S.~Heinemeyer, W.~Hollik, H.~Rzehak and G.~Weiglein,
  JHEP\ {\bf 0702}, 047  (2007)
  [hep-ph/0611326].


  \bibitem{bsg_sm}
  M.~Misiak, H.~M.~Asatrian, K.~Bieri, M.~Czakon, A.~Czarnecki, T.~Ewerth, A.~Ferroglia and P.~Gambino {\it et al.},
  Phys.\ Rev.\ Lett.\  {\bf 98}, 022002 (2007)
  [hep-ph/0609232];
  M.~Misiak, H.~M.~Asatrian, R.~Boughezal, M.~Czakon, T.~Ewerth, A.~Ferroglia, P.~Fiedler and P.~Gambino {\it et al.},
  arXiv:1503.01789 [hep-ph].

  \bibitem{bsg_hfag}
Heavy Flavor Averaging Group, http://www.slac.stanford.edu/xorg/hfag/


\bibitem{superiso}
 F.~Mahmoudi,
  Comput.\ Phys.\ Commun.\  {\bf 178}, 745 (2008)
  [arXiv:0710.2067 [hep-ph]];
  Comput.\ Phys.\ Commun.\  {\bf 180}, 1579 (2009)
  [arXiv:0808.3144 [hep-ph]].

\bibitem{bsmm}
V.~Khachatryan {\it et al.}  [CMS and LHCb Collaborations],
  Nature (2015)
  [arXiv:1411.4413 [hep-ex]].
  
\bibitem{top_mass_th} 
  [ATLAS and CDF and CMS and D0 Collaborations],
  arXiv:1403.4427 [hep-ex].
  
    

  
  \bibitem{lhc_future}
  The ATLAS Collaboration, ATLAS NOTE, ATL-PHYS-PUB-2014-010.
  
  
  \bibitem{ntp_dm} 
  R.~Allahverdi and M.~Drees,
  Phys.\ Rev.\ Lett.\  {\bf 89}, 091302 (2002)
  [hep-ph/0203118];
R.~Allahverdi and M.~Drees,
  Phys.\ Rev.\ D {\bf 66}, 063513 (2002)
  [hep-ph/0205246].
 G.~B.~Gelmini and P.~Gondolo,
  Phys.\ Rev.\ D {\bf 74}, 023510 (2006)
  [hep-ph/0602230];
  Y.~Kurata and N.~Maekawa,
  Prog.\ Theor.\ Phys.\  {\bf 127}, 657 (2012)
  [arXiv:1201.3696 [hep-ph]].

\bibitem{ntp_dm2} 
  K.~Harigaya, M.~Kawasaki, K.~Mukaida and M.~Yamada,
  Phys.\ Rev.\ D {\bf 89}, no. 8, 083532 (2014)
  [arXiv:1402.2846 [hep-ph]].

  
  \bibitem{relic_obs}
  G.~Hinshaw {\it et al.}  [WMAP Collaboration],
  Astrophys.\ J.\ Suppl.\  {\bf 208}, 19 (2013)
  [arXiv:1212.5226 [astro-ph.CO]];
  P.~A.~R.~Ade {\it et al.}  [Planck Collaboration],
  Astron.\ Astrophys.\  {\bf 571}, A16 (2014)
  [arXiv:1303.5076 [astro-ph.CO]].
  
  
  \bibitem{coan}
   K.~Griest and D.~Seckel,
  Phys.\ Rev.\ D {\bf 43}, 3191 (1991).
  
\bibitem{mirage1} 
K.~Choi, A.~Falkowski, H.~P.~Nilles, M.~Olechowski and S.~Pokorski,
  JHEP {\bf 0411}, 076 (2004)
  [hep-th/0411066];
  K.~Choi, A.~Falkowski, H.~P.~Nilles and M.~Olechowski,
  Nucl.\ Phys.\ B {\bf 718}, 113 (2005)
  [hep-th/0503216].
  
  \bibitem{endo} 
  M.~Endo, M.~Yamaguchi and K.~Yoshioka,
  Phys.\ Rev.\ D {\bf 72}, 015004 (2005)
  [hep-ph/0504036].
  
  \bibitem{ccb}
  J.~M.~Frere, D.~R.~T.~Jones and S.~Raby,
  Nucl.\ Phys.\ B {\bf 222}, 11 (1983);
   J.~F.~Gunion, H.~E.~Haber and M.~Sher,
  Nucl.\ Phys.\ B {\bf 306} (1988) 1;
  J.~A.~Casas, A.~Lleyda and C.~Munoz,
  Nucl.\ Phys.\ B {\bf 471}, 3 (1996)
  [hep-ph/9507294];
  A.~Kusenko, P.~Langacker and G.~Segre,
  Phys.\ Rev.\ D {\bf 54}, 5824 (1996)
  [hep-ph/9602414].
  
  \bibitem{ccb_recent} 
   J.~E.~Camargo-Molina, B.~O'Leary, W.~Porod and F.~Staub,
  JHEP {\bf 1312}, 103 (2013)
  [arXiv:1309.7212 [hep-ph]];
  D.~Chowdhury, R.~M.~Godbole, K.~A.~Mohan and S.~K.~Vempati,
  JHEP {\bf 1402}, 110 (2014)
  [arXiv:1310.1932 [hep-ph]];
  N.~Blinov and D.~E.~Morrissey,
  JHEP {\bf 1403}, 106 (2014)
  [arXiv:1310.4174 [hep-ph]];
  J.~E.~Camargo-Molina, B.~Garbrecht, B.~O'Leary, W.~Porod and F.~Staub,
  Phys.\ Lett.\ B {\bf 737}, 156 (2014)
  [arXiv:1405.7376 [hep-ph]].
  
  
  
  \bibitem{moduli1}
  M.~Endo, K.~Hamaguchi and F.~Takahashi,
  Phys.\ Rev.\ Lett.\  {\bf 96}, 211301 (2006)
  [hep-ph/0602061]; 
   S.~Nakamura and M.~Yamaguchi,
  Phys.\ Lett.\ B {\bf 638}, 389 (2006)
  [hep-ph/0602081].
  
\bibitem{linde} 
  A.~D.~Linde,
  Phys.\ Rev.\ D {\bf 53}, 4129 (1996)
  [hep-th/9601083].

\bibitem{ty_nty} 
  F.~Takahashi and T.~T.~Yanagida,
  JHEP {\bf 1101}, 139 (2011)
  [arXiv:1012.3227 [hep-ph]];
  Phys.\ Lett.\ B {\bf 698}, 408 (2011)
  [arXiv:1101.0867 [hep-ph]];
  K.~Nakayama, F.~Takahashi and T.~T.~Yanagida,
  Phys.\ Rev.\ D {\bf 84}, 123523 (2011)
  [arXiv:1109.2073 [hep-ph]];
  Phys.\ Rev.\ D {\bf 86}, 043507 (2012)
  [arXiv:1112.0418 [hep-ph]];
  Phys.\ Lett.\ B {\bf 714}, 256 (2012)
  [arXiv:1203.2085 [hep-ph]].

\bibitem{kklt_general}
   K.~Choi, K.~S.~Jeong and K.~i.~Okumura,
  JHEP {\bf 0509}, 039 (2005)
  [hep-ph/0504037].

\end{thebibliography}
\end{document}